\documentclass[12pt,prb, amssymb, amsmath, aps]{revtex4}
\usepackage{graphicx}

\usepackage{bm}

\newcommand{\arctanh}{\text{arctanh}}
\newcommand{\bigarrow}{\to}

\begin{document}

\title{Relativistic velocity space, Wigner rotation and Thomas
precession}
\author{John A. Rhodes}
\email{jrhodes@bates.edu}
\affiliation{Department of Mathematics, Bates College, Lewiston, Maine
04240}

\author{Mark D. Semon}
\email{msemon@bates.edu}
\affiliation{Department of Physics and Astronomy, Bates College,
Lewiston, Maine 04240}

\date{September 21, 2003}

\begin{abstract}
We develop a relativistic velocity space called \emph{rapidity space}
from the single assumption of Lorentz invariance, and use it to
visualize and calculate effects resulting from the successive
application of non-colinear Lorentz boosts. In particular, we show how
rapidity space provides a geometric approach to Wigner rotation and 
Thomas precession in the same way that spacetime provides a geometrical
approach to kinematic effects in special relativity.
\end{abstract}

\maketitle
\section{Introduction}\label{S1}

The most commonly used Lorentz transformation relates an inertial frame $S$ to an inertial frame $S'$ moving with a velocity $\vec{v}$ along the $x$-axis of $S$.  If this Lorentz transformation also preserves the orientation of the spatial axes of $S$ and leaves the sign of its time component unchanged (as is usually the case), then it is called a \emph{boost}.  In some cases an inertial frame $S'$ is obtained from an inertial frame $S$ by \emph{two successive} boosts.  If the two successive boosts are non-colinear then, contrary to what one might expect, the single Lorentz transformation that is their resultant is not a pure boost but rather is the product of a boost and a rotation. The unexpected rotation was discovered by Thomas\cite{Thomas} in 1926, and derived thirteen years later by Wigner\cite{Wigner} in his seminal article on representations of the Lorentz group. If successive non-colinear boosts return the spatial origin of $S'$ to the spatial origin of $S$, then all of the Thomas-Wigner rotations along the way combine to produce a net rotation of $S'$ with respect to $S$ called the \emph{Thomas precession.}\cite{Fisher,Jackson,Sard,Arzelies,Goldstein,Kroemer}

Thomas precession is an essential part of quantum courses discussing
relativistic corrections to the Hamiltonian of a hydrogen atom because
it changes the non-relativistic form of the spin-orbit term by a factor
of one-half. Rather than derive this result, however, some texts
state it without giving any references,\cite{Gasiorowicz} some
state it and reference only Thomas' original article,\cite{Liboff}
while others state it and appeal to the Dirac equation for its
justification.\cite{Shankar,Sakurai}

The few texts and journal articles which do derive Thomas precession often use mathematics that is somewhat sophisticated,\cite{Costella1} such as infinitesimal generators of the Lorentz
group,\cite{Jackson} ``a weakly associative-commutative
groupoid,''\cite{Ungar}
``gyrogroups and gyrovector
spaces,''\cite{Ungar1} the Gibbs method
for adding finite rotations,\cite{Ben-Menahem}
holonomy group transformations and Clifford-Dirac
algebra,\cite{Urbantke} the tetrad
formalism,\cite{Goedecke} Fermi-Walker
transport,\cite{Hamilton} or unboosted Fermi-Walker
frames,\cite{Rowe}.  Although several texts do present straightforward algebraic derivations, they are long and somewhat tedious.\cite{Sard,Arzelies} Recalling that many relativistic effects are easier to derive and understand when treated geometrically in spacetime, we wondered if there was a relativistic velocity space in which Thomas precession (and other effects involving successive non-colinear boosts) could be treated geometrically and in this way also made easier to understand.

Perhaps the most intriguing approach to constructing a relativistic
velocity space was mentioned in the 1950's by Landau and
Lifshitz.\cite{Landau1} They begin an exercise for the reader by
noting that given two non-colinear relativistic velocities $\vec{v}$
and $\vec{v}+ d\vec{v}$, the relative velocity $d\vec{v}$ can be
considered as a line element in a three-dimensional velocity space
in which each point is specified by the azimuthal and polar angles of
$\vec{v}$ and a radial coordinate equal to a function of $v$ called the
\emph{rapidity}. Landau and Lifshitz then ask the reader to show that this relativistic velocity space is non-Euclidean, with a hyperbolic geometry.

Landau and Lifshitz do not reference the origin of this exercise, so it
is not clear if they discovered the velocity space themselves
or are drawing upon work published previously in the Russian
literature. Both Pauli\cite{Pauli} and Rosenfeld\cite{Rosenfeld} credit a paper written in
Russian by the Croatian mathematician Vari\c{c}ak as the first place in
which relativistic velocity addition was related to the analog of
vector addition in a hyperbolic space. Pauli cites four additional
articles (also written in Russian) by Vari\c{c}ak, published between
1910 and 1919, and Rosenfeld notes that Vari\c{c}ak summarized and
expanded upon his work in a book (written in Russian) published in
1924.  Rosenfeld cites only two references to work on this subject that appeared after 1924, both of which are books written in Russian.  One was published in 1963 and the other in 1965.

Given this history, it seems likely that Landau and Lifshitz's text was
the first written in English to mention a relativistic velocity space
with hyperbolic geometry. Indeed, in 1997, when 
Aravind\cite{Aravind} showed how the Thomas-Wigner
rotation and Thomas precession had properties identical to those of
areas in a hyperbolic space, he credited this discovery to ``the crucial hint \ldots from Landau and Lifshitz\ldots .'' More recently, Criado and Alamo\cite{Criado} chose a hyperboloid in spacetime to represent a relativistic velocity space, which they mapped onto a unit disk with hyperbolic geometry (called the \emph{Poincar\'e disk}). They then
drew on results from non-Euclidean geometry, such as the law of cosines
and the equations of geodesics in a hyperbolic space, to show how
certain properties of hyperbolic triangles correspond to certain
properties of relativistic velocities and velocity addition.

The interesting results in Refs.~\onlinecite{Aravind} and
\onlinecite{Criado} are not readily accessible to many physicists because they assume a familiarity with formulas and theorems from non-Euclidean geometry. Furthermore,
although these articles make the connection between relativistic
velocity addition and hyperbolic geometry compelling, neither
explains this connection nor develops it systematically from first
principles.

The purpose of this paper is to derive a relativistic velocity space
(called \emph{rapidity space}) from first principles, and to demonstrate how
it provides a geometric approach to solving problems involving the
relativistic addition of non-colinear velocities and successive,
non-colinear, Lorentz boosts. The development is 
self-contained and assumes no previous knowledge of hyperbolic
geometry. Beginning with the single requirement of Lorentz
invariance, we construct rapidity space using an approach that
parallels the one used to establish spacetime. We find
that just as many kinematic effects in special relativity are more
easily and elegantly understood once the spacetime metric
\begin{equation}
ds^2=dx^2 + dy^2 -
c^2 dt^2\label{E2}
\end{equation}
is established (in a spacetime with two
spatial and one time dimension), so too are many aspects of the addition
of non-colinear boosts more easily understood once the rapidity space
metric
\begin{equation}
ds^2=\left(\dfrac{2}{1-x^2-y^2}\right) ^2(dx^2 + dy^2)\label{E3}
\end{equation}
is established (with $x$ and $y$ related to the usual components of velocity, as defined in
Sec.~\ref{S5}). In particular, once the main properties of rapidity
space have been developed, exact expressions for the Thomas-Wigner
rotation and Thomas precession can be found geometrically. 
Furthermore, working in rapidity space allows various qualitative
aspects of these effects to be deduced geometrically, some of which
are more difficult to prove with algebraic equations alone. Indeed, we
have found relativistic velocity (rapidity) space to be
as useful for understanding the relativistic addition of
non-colinear velocities and Lorentz boosts as spacetime has been for understanding kinematic effects in special relativity.

As mentioned above, this paper derives a 2D relativistic velocity space (called rapidity space) from the single assumption of Lorentz invariance.  Although developing the material in this way is logically satisfying, and has the additional benefit of unifying (and occassionally correcting) results presented previously, it does make the paper rather long.  Those who would rather bypass the derivations and proofs, and simply accept that there is a relativistic velocity space whose metric is given by Eq.~(\ref{allequations28c}) and whose geodesics are the ones described at the end of Sec.~\ref{S6}, can proceed directly to the applications presented in Sec.~\ref{S7}. Sections ~\ref{S2} through \ref{S6} present the proofs and derivations needed to establish the relativistic velocity space and its properties, while Sec.~\ref{S8} is included for those interested in how some of what is presented here is expressed using group theory, quaternions, spinors, etc.

\section{Notation and Background}\label{S2}
Consider two inertial frames $S$ and $S'$ whose origins are coincident
when $t=t'=0$, and whose $x$ and $x'$ axes are aligned. (Two inertial frames related in this way are said to be in the \emph{standard configuration}, and it is easy to show
that the linearity of Lorentz transformations always makes this choice possible for any two inertial frames $S$ and $S'$.\cite{Rindler1}) As mentioned in the Introduction, a non-trivial Lorentz
transformation from $S$ to an inertial frame $S'$ moving with a velocity
$\vec{v}$ with respect to $S$ is called a boost if it preserves the
orientation of the spatial axes and leaves the sign of the time
component unchanged. If the boost is in the $x$-direction, then the
transformation equations are
\begin{subequations}
\label{allequations1}
\begin{eqnarray}
x'&=&\gamma(x-vt),\label{equation1a} \\
y'&=&y,\label{equation1b} \\ z'&=&z,\label{equation1c} \\ t'&=&\gamma
\left(t-\dfrac{vx}{c^2}\right),\label{equation1d}
\end{eqnarray}
\end{subequations}
with $\gamma = 1/\sqrt{1-v^2/c^2}$. In what follows, we
restrict ourselves to spacetimes with one time and two spatial
dimensions because this is sufficient for understanding the most common
cases of Thomas rotation and precession.\cite{Rhodes}

For convenience, let $x_1 =x$, $x_2 = y$, $x_3 = ct$, and $\beta =
v/c$. Using this notation, $\gamma = (1-\beta^2)^{-\frac{1}{2}}$, and
Eq.~(\ref{allequations1}) becomes
\begin{equation}
\begin{pmatrix}
x_1 ' \\
x_2 ' \\ x_3 ' \end{pmatrix}
=
\begin{pmatrix}
\gamma & 0 &
-\gamma\beta \\
0 & 1 &0 \\
 -\gamma\beta & 0 & \gamma
\end{pmatrix}
\begin{pmatrix}
x_1 \\ x_2 \\ x_3 \end{pmatrix}.
\label{E5}
\end{equation}

If we let $\bm{x}$ represent the column matrix on the right-hand side of
Eq.~(\ref{E5}), then the length (norm) squared of $\bm{x}$ can be
expressed as
\begin{equation}
x_1^2 + x_2^3-x_3^2=\bm{x}^T
\begin{pmatrix}
1 &0 & 0\\ \\ 0 &1 &0 \\ \\ 0 &0 &-1
\end{pmatrix}
\bm{x}
=\bm{x}^T\bm{G}\bm{x}.
\end{equation}
More generally, any linear
transformation $\bm{\Lambda}$ is a Lorentz transformation if and only if
it leaves the spacetime metric
\begin{equation}
\bm{x'}^T\bm{G}\bm{x'}=\bm{x}^T\bm{G}\bm{x}
\end{equation}
invariant for all $\bm{x}$, or equivalently, if and only if
\begin{equation}
\bm{\Lambda}^T\bm{G}\bm{\Lambda}=\bm{G}.\label{E5a}
\end{equation}

\section{Relativistic Velocity Addition and the Rapidity}\label{S3}

The \emph{rapidity $\phi$} of a boost $\vec{\beta}$ is defined as
\begin{equation}
\phi \equiv \arctanh\,\beta .\label{E1}
\end{equation}
Thus,
\begin{subequations}
\label{allequations2}
\begin{eqnarray} \beta&=&\tanh\phi,\label{equation2a} \\
\gamma&=&\cosh\phi,\label{equation2b} \\
\gamma\beta&=&\sinh\phi.\label{equation2c}
\end{eqnarray}
\end{subequations}

Using the rapidity allows Lorentz boosts to be expressed in two
alternative and interesting ways. In the first, if we substitute the
rapidity into Eq.~(\ref{E5}), we obtain
\begin{equation}
\begin{pmatrix}
x_1 ' \\\\
x_2 '\\\\ x_3 '
\end{pmatrix}=
\begin{pmatrix}
\cosh\phi &0 &
-\sinh\phi\\\\ 0 &1 &0\\\\ -\sinh\phi & 0 & \cosh\phi
\end{pmatrix}
\begin{pmatrix}
x_1
\\\\ x_2 \\\\ x_3
\end{pmatrix} \label{E6},
\end{equation}
which
illustrates that the rapidity can be interpreted as an imaginary rotation
angle in spacetime.

A second way of expressing Lorentz boosts is found by introducing the
new coordinates $(\xi,\eta)$, with\cite{Rindler2}
\begin{equation}
\xi \equiv x_3 + x_1 \quad \text{and} \quad \eta \equiv
x_3 - x_1.
\end{equation}
If we use these coordinates, Eq.~(\ref{E6}) can be
expressed in the simple form
\begin{subequations} \label{allequations3}
\begin{eqnarray} \xi'&=& e^{-\phi}\xi \\ \label{allequations3a}
\eta'&=& e^{\phi}\eta \label{allequations3b}.
\end{eqnarray}
\end{subequations} For future reference, note that
Eq.~({\ref{allequations3}) also can be written in the form
\begin{subequations}
\label{allequations4}
\begin{eqnarray}
\xi'&=&(\gamma -\gamma\beta)\,\xi =\sqrt{\dfrac{1-\beta}{1+\beta}}\;\xi \\
\label{allequations4a} \eta'&=&(\gamma
+\gamma\beta)\,\eta=\sqrt{\dfrac{1+\beta}{1-\beta}}\;\eta
\label{allequations4b}.
\end{eqnarray}
\end{subequations}
The $\xi$ and
$\eta$ coordinate axes lie on the light cone ($\sqrt{x^2+y^2}=\pm ct$)
and transform into themselves under this type of Lorentz boost. 
Expressed in another way, these axes are eigenvectors of the boost in
Eq.~(\ref{E6}) with real eigenvalues $e^{\pm\phi}$ which, as can be seen from Eq.~(\ref{allequations3}) and ~(\ref{allequations4}), are simply the blue- and redshift factors
in the relativistic Doppler effect.\cite{Ref1}

The rapidity is most commonly used to simplify the addition of colinear
relativistic velocities. As is well known, the relativistic addition of
two colinear velocities $\vec{v}_1$ and $\vec{v}_2$ gives a resultant
(colinear) velocity $\vec{v}$ with magnitude
\begin{equation}
v =
\dfrac{v_1 + v_2}{1+(v_1 v_2/c^2)}\iff \beta = \dfrac{\beta_1 +
\beta_2}{1+\beta_1 \beta_2} \label{E7}.
\end{equation}
The correct generalization of Eq.~(\ref{E7}) to the relativistic addition of $n$ colinear velocities is not obvious. However, if we reexpress Eq.~(\ref{E7}) using the rapidity, we
find that
\begin{equation}
 \phi=\arctanh\,\beta = \arctanh\left(\dfrac{\beta_1 +
\beta_2}{1+\beta_1 \beta_2}\right) .
\end{equation}
Using the identity
\begin{equation}
\arctanh\,\alpha =
\dfrac{1}{2}\ln\left(\dfrac{1+\alpha}{1-\alpha}\right),\label{E8}
\end{equation}
we have
\begin{equation}
\phi
=\dfrac{1}{2}\ln\dfrac{(1+\beta_1)(1+\beta_2)}{(1-\beta_1)(1-\beta_2)}.
\end{equation}
Thus,
\begin{subequations}
\label{allequations5}
\begin{eqnarray} \phi
&=&\dfrac{1}{2}\ln\left(\dfrac{1+\beta_1}{1-\beta_1}\right) +
\dfrac{1}{2}\ln\left(\dfrac{1+\beta_2}{1-\beta_2}\right)
\label{allequations5a}\\ &=&\arctanh\,\beta_1 +
\arctanh\,\beta_2
\label{allequations5b}\\
&=& \phi_1 + \phi_2
\label{allequations5c}.
\end{eqnarray} \end{subequations}

If we express the sum of two colinear velocities in the form of 
of Eq.~(\ref{allequations5}) rather than in
the usual form of Eq.~(\ref{E7}), the relativistic sum of $n$ colinear
velocities
$\vec{\beta_1}, \vec{\beta_2}, \ldots, \vec{\beta_{n}}$ is easily shown
to have magnitude $\beta$, with
\begin{subequations}
\label{allequations6}
\begin{eqnarray}
\phi
&=&\arctanh\,\beta \\
&=&\dfrac{1}{2}\ln\dfrac{(1+\beta_1)(1+\beta_2)\ldots(1+\beta_n)}{(1-\beta_1)(1-\beta_2)\ldots(1-\beta_n)}\label{allequations6b}\\
&=&\frac{1}{2}\ln\left(\dfrac{1+\beta_1}{1-\beta_1}\right)+\ldots+\frac{1}{2}\ln\left(\dfrac{1+\beta_n}{1-\beta_n}\right)\label{allequations6c}\\
&=&\arctanh\,\beta_1 + \dots+ \arctanh\,\beta_n \label{allequations6d}\\
&=&\phi_1 + \phi_2 + \ldots + \phi_n \label{allequations6e}.
\end{eqnarray}
\end{subequations}
Thus, the rapidity provides an easy way to express the relativistic sum of $n$ colinear velocities when $n\geq 2$.

Note that using the Lorentz transformation Eq.~(\ref{allequations3}) allows us to express
Eq.~(\ref{allequations6}) in another useful
form.\cite{Parker} If we combine
Eqs.~(\ref{allequations3b}) and (\ref{allequations4b}), we have
\begin{equation}
e^{\phi} = \sqrt{\dfrac{1+\beta}{1-\beta}} . \label{E10}
\end{equation}
Using this relation in Eq.~(\ref{allequations6e}) we find that
\begin{equation}
e^{\phi}=e^{\phi_1}e^{\phi_2}\ldots
e^{\phi_n},
\end{equation}
which implies that
\begin{equation}
\left(\dfrac{1+\beta}{1-\beta}\right)=
\left(\dfrac{1+\beta_1}{1-\beta_1}\right)\left(\dfrac{1+\beta_2}
{1-\beta_2}\right)\ldots\left(\dfrac{1+\beta_n}{1-\beta_n}\right).
\label{E9}
\end{equation}
Equation~(\ref{E9}) provides a surprisingly
easy way to find the resultant $\beta$ of the relativistic sum of $n$
colinear boosts.

\section{Rapidity Space}\label{S4}

In this section we derive a 2D relativistic velocity space, called
rapidity space, directly from the 3D spacetime of special relativity
(that is, from the spacetime with two spatial and one time dimension). 
If we use the coordinates $x_1$, $x_2$, and $x_3$ defined in
Sec.~\ref{S2}, the square of the line element for this spacetime is
\begin{equation}
ds^2 = (dx_1)^2 + (dx_2)^2 - (dx_3)^2.\label{E9b}
\end{equation}
We choose this particular form of $ds^2$ because in the
$x_3=0$ plane, it reduces to the usual Euclidean relation
\begin{equation}
ds_E^2 = (dx_1)^2 + (dx_2)^2. \end{equation}

Suppose we fix ourselves in one inertial frame and consider another with
the same spacetime origin but moving with a velocity $\vec{v}$ relative
to the first. The spatial origin of this second frame appears to us as
following a straight line $(x(t),y(t))= \vec v t,$ where $(v/c) <1$.
Thus its trajectory is a line emanating from the origin and lying within
the light cone. This line also can be described in $x_i$-coordinates
as the one formed by all the scalar multiples of the vector $(\vec
\beta, 1)$, where $\vec \beta=\vec v/c$. If we turn this statement
around, we can say that every straight line through the origin that
lies within the light cone represents the trajectory of the origin of
some inertial frame traveling with a velocity $\vec{v}$ relative to the
fixed inertial frame represented by the spacetime.

Another way of describing all the straight lines through the origin and
within the light cone is to note that each can be viewed as the $x'_3$
axis of some inertial frame obtained from the original $(x_1, x_2, x_3)$
frame by a unique boost. Because each $x'_3$ axis corresponds to one
particular velocity (and vice versa), we can create a model of velocity
space by choosing one point from each $x'_3$ axis. The set of all such
points is a velocity space because each point in it represents
a unique velocity, and because all velocities $\vec \beta$ with
magnitude $\beta <1$ will be represented.

There are several natural ways to construct a relativistic velocity space from the set of points defined in the previous paragraph. For example, we could start with all the points lying in the plane $x_3=1$ and inside the lightcone; alternatively, we could start with all the points lying on the hyperboloid with $x_3 > 0$.

%\begin{figure}
%\begin{center}
%\includegraphics[width=3.5in, height=3in]{rhodes1.eps}
%\end{center}
%\caption{The
%hyperboloid and Klein (or simultaneity) models of rapidity space. Note
%that any possible time-axis intersects each surface exactly once, so the
%point of intersection can be used to represent that axis. Part of the
%light cone has been removed to make the figure easier to
%understand.} \label{F1} 
%\end{figure}

As shown in Fig. the first choice is the \emph{simultaneity
plane} $x_3=1$ for an observer in the inertial frame represented by our
spacetime, while the second is the set of all points for which the proper
time $\tau=1$. The first choice results in a velocity space known as
the
Klein model, which is not the best choice for our purposes because
angles in this model do not appear like Euclidean angles (that is, the
Klein model is not conformal.\cite{ftnote3}}) The reason we would like
a conformal model is that the relativistic addition of non-colinear
velocities, the Thomas-Wigner rotation and the Thomas
precession, are built on understanding angles between successive
boosts, so only those spaces in which angles behave like Euclidean
angles can be expected to offer the geometric insight we seek.

The second choice, of all the points that lie on the hyperboloid of
revolution, does result in a conformal velocity space and is in fact the
one chosen in Ref.~\onlinecite{Criado}. This choice is
both reasonable and convenient because the metric on this space and the
geometric properties that follow from it are well-known to
mathematicians. However, few of us are good at judging angles on a
curved surface.

So how do we motivate or justify choosing one model over another? Given
that we are free to choose any surface created by any method of
choosing one point from each $x'_3$ axis, why choose the hyperboloid?
How do we know there isn't some other way of choosing a point from each
$x'_3$ axis that will lead to an even more convenient or appropriate
model of velocity space?

Rather than trying to justify one choice over another after the fact, we
construct a model of relativistic velocity space from first principles
by following the method used to derive spacetime and the spacetime
metric. In that case, spacetime is developed from the physical
requirement that the speed of light is independent of the motion of the
source and is the same in all inertial frames. The mathematical
statement of this property, $r^2=(ct)^2$, implies that
\begin{equation}
x^2 + y^2 - (ct)^2 = 0,
\end{equation}
and leads to interpreting the quantity
\begin{equation}
x^2 + y^2 - (ct)^2
\end{equation}
as the square of a distance in a 3D spacetime where Lorentz
transformations are represented by linear coordinate transformations
$\bm{\Lambda}$ satisfying Eq.~(\ref{E5a}). Thus, rather that choosing
\emph{a priori} the nature of spacetime, a physical invariance is used
to deduce a metric that determines its mathematical properties.

We use this same approach to deduce the geometry of a relativistic
velocity space. We note that Lorentz transformations acting on spacetime also act on the set of rays inside the light cone emanating from the origin, and that each of these rays
has a one-to-one correspondence with a rapidity. Thus, the invariance
of the spacetime metric Eq.~(\ref{E9b}) under Lorentz transformations
can be used to define a metric on the rays (or rapidities) that also is
Lorentz invariant.

To find this metric, first note that because we are choosing the points in
rapidity space to correspond to rays inside the light cone that emanate
from the spacetime origin, the metric in rapidity space should be
expressible in terms of the spacetime coordinates. That is, there should
be functions $f_{i,j}$ such that the line element squared in rapidity
space can be expressed in the form
\begin{equation}
ds^2=\sum_{i,j=1}^3f_{i,j}(x_1,x_2,x_3)dx_i dx_j.
\end{equation}
However,
because any two points on the same ray in spacetime specify the same
rapidity, the line element in rapidity space must be the same regardless
of which spacetime points on the ray we choose. Thus, for any $\lambda$,
we require
\begin{equation}
ds^2(x_1,x_2,x_3)=ds^2(\lambda x_1,\lambda
x_2, \lambda x_3).\label{E9c}
\end{equation}

The spacetime form $ds^2=dx_1^2+dx_2^2-dx_3^2$ does not have this
property because
\begin{equation}
ds^2(\lambda x_1,\lambda x_2, \lambda
x_3)=\lambda^2 ds^2(x_1,x_2,x_3).\label{E9d} \end{equation}
However, we can obtain a $ds^2$ with the property given in
Eq.~(\ref{E9c}) by using a simple but clever trick: first take the
logarithm of both sides of Eq.~(\ref{E9d}) (which changes the
multiplication by $\lambda^2$ into the addition of $\ln\lambda^2$), and
then differentiate (so that the $\ln\lambda^2$ term disappears).

More formally, the spacetime inner product,
\begin{equation}
q(x,y) =-x\cdot y= -x_1y_1-x_2y_2+x_3y_3,
\end{equation}
is positive for rays within the light cone, and has the
property
\begin{equation}
q(\lambda_1 x,\lambda_2 y) =
\lambda_1\lambda_2 q(x,y).
\end{equation}
By taking the
logarithm of both sides, we find
\begin{equation}
\ln{q(\lambda_1 x,\lambda_2 y)}=\ln{\lambda_1}+\ln{\lambda_2}+\ln{q(x,y)}.
\end{equation}
Finally, taking the differential of both sides respect to $x$ and $y$, we obtain
\begin{equation}
d_x d_y
\ln{q(\lambda_1 x,\lambda_2 y)}=d_x d_y
\ln{q(x,y)}.
\end{equation}
In this way we are led to look for a rapidity space metric whose inner
product has the form
\begin{align}
d_{x}d_{y}\ln{(-x\cdot y)}&=
d_{x}d_{y}[\ln{(x_3y_3 - x_1y_1 - x_2y_2)}] \\ &=
d_{x}\left[\frac{x_3dy_3 - x_1dy_1 - x_2dy_2}{x_3y_3 - x_1y_1 -
x_2y_2}\right] \\ &= (x\cdot y)^{-2}[(dx\cdot dy)(x\cdot y) - (x\cdot
dy)(y\cdot dx)].
\end{align}
Setting $x=y$ we see that the line element squared in this relativistic velocity space should
have the form
\begin{equation}
ds^2=K(x\cdot x)^{-2}\big[(dx\cdot dx)(x\cdot x) - (x\cdot
dx)^{2}\big],\label{E13}
\end{equation}
where $K$ is an arbitrary constant.

Recall that even though the line element in Eq.~(\ref{E13}) is expressed
in terms of spacetime coordinates, it also is a function of the rays
inside the light cone and through the origin on which those spacetime
points lie.

As mentioned above, in order to better visualize the geometry that
follows from $ds^2$ in Eq.~(\ref{E13}), we may choose as a model of
relativistic velocity space any surface that intersects each ray in
exactly one point. If such a surface is expressed as
\begin{equation}
x_3=g(x_1,x_2),
\end{equation}
then by substituting this expression for
$x_3$ into Eq.~(\ref{E13}), we can express the metric in terms of the
two coordinates $x_1$ and $x_2$.

Although we have great freedom in selecting the function $g$, several judicious choices will greatly simplify our model. First, because Lorentz transformations include
rotations, and because we seek a metric that is invariant under Lorentz
transformations, it is natural to choose a surface that has rotational
symmetry. Thus, we require that
\begin{equation}
x_3=g\left(\sqrt{(x_1)^2 + (x_2)^2}\right)\equiv g(r).\label{E14}
\end{equation}
To find $dx_3$ in terms of $x_1$ and
$x_2$, we apply the chain rule to Eq.~(\ref{E14}), giving
\begin{equation}
dx_3=\dfrac{g'}{r}\left(x_1dx_1 +
x_2dx_2\right),\label{E15}
\end{equation}
with
\begin{equation}
g'\equiv\dfrac{dg}{dr}.
\end{equation}
Because
\begin{equation}
-x\cdot x =
x_3^2-x_1^2- x_2^2 = g^2-r^2,
\end{equation}
we can rewrite
Eq.~(\ref{E13}) as
\begin{equation}
\begin{split} 
&ds^2=
K(g^2-r^2)^{-2}\Bigl[(dx_3^2 - dx_1^2 - dx_2^2)(g^2-r^2)\\
&\qquad\quad-(x_3dx_3-x_1dx_1-x_2dx_2)^2\Bigr].\label{E15a}
\end{split}
\end{equation}
If we use Eqs.~(\ref{E14}) and ~(\ref{E15}) in
Eq.~(\ref{E15a}) and do some algebra, we find that
\begin{equation}
ds^2 = \left[\dfrac{K}{r^2 - g^2}\right](dx_1^2 + dx_2^2) -
K\left[\dfrac{{g'}^2(r^2 - g^2)}{r^2} +
\left(1-\dfrac{gg'}{r}\right)^2\right]\left(\dfrac{x_1dx_1
+x_2dx_2}{r^2-g^2}\right)^2. \label{E16}
\end{equation}

Although there is still much freedom in our choice of the surface $g$,
we now impose our desire to have a model that is
conformal.\cite{ftnote3a} We note that a metric will
be conformal if it is a multiple of the Euclidean metric, even if the
multiplicative factor varies from point to point. Thus, Eq.~(\ref{E16})
will be a conformal metric if cross terms like $dx_1dx_2$ are not
present. To this end, we look for a surface $g$ for which
\begin{subequations}
\begin{align}
\dfrac{{g'}^2(r^2 - g^2)}{r^2} + \Bigl(1-\dfrac{gg'}{r}\Bigr)^2&=0, \\
\noalign{\noindent which implies that}
g'(g'r-2g)+r&=0. \label{E17}
\end{align}
\end{subequations}

One way to solve this first-order non-linear differential equation for
$g$ is to look for solutions of the form
\begin{equation}
g = Ar^2+Br+C.
\label{E18}
\end{equation}
If we substitute thus form for $g$ into
Eq.~(\ref{E17}), we find that $g$ will be a solution if the coefficients
$A$, $B$ and $C$ satisfy the conditions
\begin{subequations}
\begin{align}
2AB &=0, \\
B^2 + 4AC&=1, \\ 2BC &=0.
\end{align}
\end{subequations}

One set of coefficients that satisfies these conditions is $A=\pm
\frac{1}{2}$, $C=\pm \frac{1}{2}$, and $B=0$, in which
case\cite{ftnote4}
\begin{equation}
g =
\pm\Bigl(\dfrac{1+r^2}{2}\Bigr).\label{E19}
\end{equation}
Therefore, we can use Eq.~(\ref{E19}) in Eq.~(\ref{E16}) and find
that
\begin{subequations}
\label{allequations28}
\begin{align}
ds^2&=\Bigl(\dfrac{K}{r^2-g^2}\Bigr)(dx_1^2+dx_2^2)
\label{allequations28a} \\
&=\Bigl(\dfrac{4K}{2r^2-r^4-1}\Bigr)(dx_1^2+dx_2^2)\label{allequations28b}
\\
\noalign{\noindent which implies that}
ds^2 &=
\Bigl(\dfrac{2}{1-r^2}\Bigr)^2(dx_1^2+dx_2^2).\label{allequations28c}
\end{align}
\end{subequations}
Note that because we prefer distances in velocity space to be
non-negative and real, we have chosen $K=-1$ in
Eq.~(\ref{allequations28c}).

The surface described by Eq.~(\ref{E19}) is
\begin{equation}
x_3\equiv
g =
\dfrac{1+r^2}{2}=\dfrac{1+x_1^2+x_2^2}{2},\label{E29k}
\end{equation}
which is a paraboloid of revolution about the $x_3$ axis with vertex at
$x_1=x_2=0$ and $x_3=1/2$. The paraboloid also goes through
points with $r^2=x_1^2+x_2^2=1$ and $x_3=1$, which are on the
light cone. Not only does the paraboloid touch the light cone
at $r=1$, but the light cone is tangent to the paraboloid at this point
because
\begin{equation}
\dfrac{\Delta(ct)}{\Delta(r)}\Big|_{r=1}\bigarrow\dfrac{d}{dr}(ct)
\Big|_{r=1}=\dfrac{d}{dr}\big(\dfrac{1+r^2}{2}\big)\Bigg|_{r=1}= 1 ,
\end{equation}
which is exactly the slope of the light cone.

%\begin{figure}
%\begin{center}
%\includegraphics[width=3.5in, height=3in]{rhodes2.eps}
%\end{center}
%\caption{The
%paraboloid model. Because each non-vertical time-axis intersects the
%paraboloid in two points, the lower point is chosen as its representation
%in this model.}\label{F2}
%\end{figure}

As shown in Fig., each ray emanating from the origin and inside the light cone (with the exception of the $x_3$-axis) intersects the
paraboloid twice: once below the disk $x_3=1$ and once above it. Because
we only need a single point from each ray to create a velocity space, we
only use that part of the paraboloid that is below the disk (that is,
points on the paraboloid with $x_3<1$), as shown in Fig..

%\begin{figure}
%\begin{center}
%\includegraphics[width=3.5in, height=3in]{rhodes3.eps}
%\end{center}
%\caption{Possible
%time axes correspond to their unique points of intersection with the
%lower part of the paraboloid, which then correspond by downward
%orthogonal projection to unique points in the Poincar\'e
%disk.}\label{F3}
%\end{figure}

Finally, note that the line element squared
given in Eq.~(\ref{allequations28c}) can be expressed in plane polar
coordinates as
\begin{equation}
ds^2=\Bigl(\dfrac{2}{1-r^2}\Bigr)^2(dr^2+r^2d\theta^2).\label{E22a}
\end{equation}

\section{Velocity Space, Rapidity Space and the Poincar\'e
Disk}\label{S5}

Although the paraboloid of revolution derived in Sec.~\ref{S4} is a
valid model of relativistic velocity space, in most cases it is much
easier to work in the space obtained by projecting this paraboloid
onto the $(x_1,x_2)$ plane by the projection $(x_1,x_2,x_3) \mapsto (x_1,x_2)$
shown in Fig..

The space created by this projection is a unit disk with the metric in
Eq.~(\ref{allequations28c}), and is known to mathematicians as the
\emph{Poincar\'e disk}. Although both the Poincar\'e and hyperboloid
models are conformal (the first in two dimensions and the second in
three), the Poincar\'e model is superior for building intuition about
the Thomas-Wigner rotation and Thomas precession because it can be
drawn in two dimensions, which makes line segments and angles easier to
visualize.

As we shall see, the distance from the origin to any point on the
Poincar\'e disk (as determined by the line element in
Eq.~(\ref{allequations28c})) is just the rapidity associated with that
point, which is why we refer to this disk as rapidity
space.\cite{ftnote12} Because points on the edge of the
disk are defined by the projection of points on the intersection of the
paraboloid and the light cone, they represent velocities with speed
$v=c$. We shall see that these points are an infinite (Poincar\'e)
distance away from any point inside the disk, reflecting the fact that
speeds can approach but never reach the speed of light.

To simplify the notation, we rename the coordinates on the disk $x\,
(\equiv x_1)$ and $y\,(\equiv x_2)$. We identify the physical
significance of the distance $s$ of any point from the origin by
evaluating\cite{ftnote5}
\begin{subequations}
\begin{align}
s
&=\!\int_{0}^{R}\!{ds}=\!\int_{0}^{R}\!\Bigl(\dfrac{2}{1-r^2}\Bigr)
\sqrt{dx^2 + dy^2} \\
&=\!\int_{0}^{R}\!\Bigl(\dfrac{2}{1-r^2}\Bigr)\sqrt{dr^2 + r^2d\theta^2}
\\ &=\!\int_{0}^{R}\!\Bigl(\dfrac{2}{1-r^2}\Bigr)dr
\\ &=\ln\Bigl(\dfrac{1+R}{1-R}\Bigr),
\end{align}
\end{subequations}
which implies that
\begin{equation}
\label{last}
s=2\,\arctanh\,R.
\end{equation}
However a point on the disk whose radial coordinate is
$R=\sqrt{a^2+b^2}$ corresponds to the spacetime point
$(a,b,\frac{a^2+b^2+1}{2})$, which lies on the ray in spacetime with
slope
\begin{eqnarray}
\dfrac{ct}{R}&=&\dfrac{(1+R^2)/2}{R}, \\
\noalign{\noindent which implies that}
\dfrac{c}{v}&=&\dfrac{1+R^2}{2R} \\
\beta
&=&\dfrac{2R}{1+R^2}=\dfrac{2\tanh (s/2)}{1+\tanh (s/2)\tanh (s/2)} \\
\beta &=&\tanh (s/2+s/2)=\tanh s.
\end{eqnarray}
We thus conclude
that the (Poincar\'e distance) $s$ of any point from the origin is
\begin{equation}
s=\arctanh\,\beta=\phi. \qquad \mbox{(the rapidity)}
\end{equation}

We can further clarify the nature of rapidity space by relating its
coordinates $(x,y)$ to the Euclidean velocity components $v_x$ and
$v_y$. First, note that a point $(x,y)$ on the disk corresponds to a
point $(x,y,\frac{1+r^2}{2})$ on the paraboloid, which means that
\begin{equation}
\beta_x=\Bigl(\dfrac{2}{1+r^2}\Bigr)x, \quad 
\beta_y=\Bigl(\dfrac{2}{1+r^2}\Bigr)y , \label{E21e}
\end{equation}
and
\begin{equation}
\beta=\Bigl(\dfrac{2}{1+r^2}\Bigr)r.\label{E21a}
\end{equation}

Equations~(\ref{E21e}) and (\ref{E21a}) give $\beta$ (and its components)
in terms of any radial coordinate $r$ (and its components). To find the
inverse relation (that is, the $r$ associated with a given
$\beta$), we first solve Eq.~(\ref{E21a}) for the magnitude of $r$:
\begin{align}
\beta &r^2-2r+\beta=0 \\ 
&r
=\dfrac{1-\sqrt{1-\beta^2}}{\beta} \\ 
&r =
\dfrac{\gamma-1}{\gamma\beta}.\label{E21b}
\end{align}
(We use only the
negative root in the quadratic formula because $r<1$). We substitute
Eq.~(\ref{E21b}) in Eq.~(\ref{E21e}) and use the identity,
\begin{equation}
\gamma^2-1=\dfrac{\beta^2}{1-\beta^2}=
(\gamma\beta)^2,\label{I1}
\end{equation}
to find
\begin{subequations}
\label{E21all}
\begin{align}
x&=\Bigl(\dfrac{1+r^2}{2}\Bigr)\beta_x
=\Bigl(\dfrac{\gamma}{\gamma+1}\Bigr)\beta_x \label{E21f} \\ y
&=\Bigl(\dfrac{1+r^2}{2}\Bigr)\beta_y=\Bigl(\dfrac{\gamma}{\gamma+1}\Bigr)
\beta_y,\label{E21g}
\end{align}
\end{subequations}
and
\begin{equation}
r=\Bigl(\dfrac{\gamma}{\gamma+1}\Bigr)\beta. \label{E21h}
\end{equation}

{}From Eq.~(\ref{E21all}) we see that the $x$
coordinate in rapidity space is proportional to $\beta_x$ and the $y$
coordinate is proportional to $\beta_y$. The proportionality factor in
both cases is
$\gamma /(\gamma +1)$, which goes to unity as $v \to c$ and to
1/2 as $v \bigarrow 0$. Thus, in the limit as $v
\bigarrow 0$, the line element squared in rapidity space
(Eq.~(\ref{allequations28c})) reduces to 
\begin{equation}
ds^2=(d\beta_x)^2 +
(d\beta_y)^2,
\end{equation}
which, of course, is the line element squared of the usual Euclidean
non-relativistic velocity space.

To summarize, we have shown that requiring the velocity space metric to
be invariant under Lorentz transformations leads to a model of
relativistic velocity space that can be represented either as a
paraboloid of revolution with its vertex at $x_1=x_2=0,x_3=\frac{1}{2}$
and top edge at $x_3=1$, or as a unit disk with the metric given in
Eq.~(\ref{allequations28c}). In this paper we refer
to this (Poincar\'e) disk as rapidity space because the distance from the
origin to any point on the disk is its rapidity. Note that from the
definition of rapidity, it is easy to see that any point on the
edge of the disk is infinitely far from any point in the disk.
Also, any point in rapidity space is related to the components of any velocity
by Eqs.~(\ref{E21e}) and (\ref{E21all}). The conformal
property of rapidity space can be seen explicitly by noting that if
$\arctan{(v_y/v_x)}$ is the angle made by a velocity vector $\vec{v}$
with the horizontal axis in real space, then, using 
Eq.~(\ref{E21e}), we have
\begin{equation}
\arctan\dfrac{v_y}{v_x}=\arctan\dfrac{y}{x}=\theta,
\end{equation}
which
means that the angle of $\vec{\beta}$ in real space is the same as the
angle $\theta$ in rapidity space. This property of rapidity space is what makes
it so useful for understanding the Thomas-Wigner rotation and Thomas
precession.

\section{Geodesics in Rapidity Space}\label{S6}
We need to understand one more aspect of rapidity space before we can
use it to investigate the relativistic addition of non-colinear velocities and boosts. When we boost from one inertial frame traveling with a velocity $\vec{v}_1$ to another
traveling with a velocity $\vec{v}_2$, we pass through the minimum number of velocities whose speeds are greater than $v_1$ and less than $v_2$.  The corresponding path in rapidity
space between the points representing these velocities is the shortest
one which, by definition, is the geodesic connecting them. Hence, to study successive non-colinear boosts, we need to identify the geodesics in rapidity space. We shall see that these geodesics are straight lines if the origin (the point representing zero velocity) is
one of the values taken on during the boost. In all other cases the
geodesics are not straight lines but rather are the arcs of 
circles. Following these arcs from one point in rapidity space to
another will bring out most of the interesting features of relativistic
velocity addition, successive Lorentz boosts, Thomas-Wigner rotations and Thomas precession.

We begin by showing that any geodesic that includes the origin of
rapidity space is a straight line. Our proof parallels the traditional
one showing that the shortest distance between two points in a Euclidean
plane is a straight line, and is accomplished by finding the path of
minimum distance connecting two points on the disk. The length of any
path connecting the origin and a point $(A,B)$ in rapidity space is
\begin{subequations}
\begin{align}
\int_{(0,0)}^{(A,B)}{ds}&=\!\int_{(0,0)}^{(A,B)}
\Bigl(\dfrac{2}{1-r^2}\Bigr)\sqrt{dr^2+r^2d\theta^2}
\\
&=\!\int_{0}^{R} \Bigl(\dfrac{2}{1-r^2}\Bigr)
\sqrt{1+r^2{\theta'}^2}\,dr, \label{integral}
\end{align}
\end{subequations}
with $R=\sqrt{A^2+B^2}$, and
\begin{equation}
\theta'\equiv\frac{d\theta}{dr}. \end{equation}
If we use the Euler-Lagrange equation, we know that the integral in
Eq.~(\ref{integral}) will be an extremum when
\begin{equation}
\dfrac{\partial}{\partial
\theta}\Bigl[\Bigl(\dfrac{2}{1-r^2}\Bigr)\sqrt{1+r^2{\theta'}^2}\Bigr] -
\dfrac{d}{dr}\dfrac{\partial}{\partial
\theta'}\Bigl[\Bigl(\dfrac{2}{1-r^2}\Bigr)\sqrt{1+r^2{\theta'}^2}\Bigr]=0.
\label{77}
\end{equation}
The first term on the left-hand side is zero, so Eq.~(\ref{77}) reduces
to
\begin{align}
\dfrac{d}{dr}\dfrac{\partial}{\partial \theta
'}\Bigl[\Bigl(\dfrac{2}{1-r^2}\Bigr)\sqrt{1+r^2{\theta'}^2}\Bigr] &= 0,
\\
\noalign{\noindent which implies that}
\dfrac{d}{dr}\Bigl[\Bigl(\dfrac{2}{1-r^2}
\Bigr)\dfrac{r^2\theta'}{\sqrt{1+r^2{\theta'}^2}}\Bigr] &= 0, \\
\noalign{\noindent and}
\Bigl(\dfrac{2}{1-r^2}\Bigr)\dfrac{r^2\theta'}
{\sqrt{1+r^2{\theta'}^2}}&=h(\theta').
\label{E22c} \end{align}
If we square both sides of Eq.~(\ref{E22c}) and
rearrange the terms, we find $h^2(Ar^6+Br^4+Cr^2+1)=4r^4{\theta'}^2$,
with $A$, $B$, and $C$ functions of $\theta'$. In order for this
equation to be satisfied for all $r$, including $r=0$,
$h(\theta')$ must be zero. Hence, from Eq.~(\ref{E22c}) we see that
$\theta'$ also must be zero, and thus $\theta$ is a constant whenever
$r=0$ lies on the path. Therefore, any geodesic in rapidity space that
includes the origin is a straight line, as shown in Fig..

%\begin{figure}
%\begin{center}
%\includegraphics[width=3.5in, height=3in]{rhodes4.eps}
%\end{center}
%\caption{Geodesics
%through the vertex of the paraboloid are formed by intersections of the
%paraboloid and vertical planes. These project to Euclidean-straight
%lines through the origin in the Poincar\'e model.}
%\label{F4}
%\end{figure}

Next we derive the geodesics in rapidity space that do not include the
origin. Because the rapidity space metric is Lorentz invariant by
construction, Lorentz transformations must send geodesics to geodesics. 
Thus, we can obtain a geodesic that does not include the origin by
applying the same boost to every point on a geodesic that does include
the origin. Rather than doing this directly, it is easier to obtain the
final geodesic geometrically by projecting back and forth between the
disk and the paraboloid. In this approach, we first identify the
geodesics on the paraboloid that correspond to geodesics through the
origin of the disk. We then apply the same boost to every point on one
of these geodesics on the paraboloid. Finally, we project the
Lorentz transformed geodesic on the paraboloid onto the disk and
find the equation that describes it.

We already have proven that any geodesic that includes the origin of
the disk is a straight line. If we project one of these straight
lines back up to the paraboloid, we see that it corresponds to a
parabola through the vertex of the paraboloid, as shown in
Fig.. Thus, any such (vertically oriented) parabola is a
geodesic. Equivalently, any one of these (vertically oriented)
parabolas can be regarded as the curve formed at the intersection of
the paraboloid and the plane defined by two axes
$x_3'$ and $x_3''$, each of which has a point on the paraboloids'
geodesic.

%\begin{figure}
%\begin{center}
%\includegraphics[width=3.5in, height=3in]{rhodes5.eps}
%\end{center}
%\caption{All
%geodesics on the paraboloid are formed by its intersection with planes
%through the origin. These project downward onto circular arcs, orthogonal
%to the unit disk, in the Poincar\'e model.}
%\label{F5}
%\end{figure}

If we now perform the same pure boost on every frame represented by a
point on a parabola passing through the vertex of the paraboloid, we
obtain a new geodesic that does not include the vertex. Because boosts
are linear transformations, planes through the origin transform into
other planes through the origin. Thus, the new geodesic can be
described as the curve created by the intersection of the paraboloid
and the new plane formed by the boosted $x_3'$ and $x_3''$ axes, as
shown in Fig.. If we project this curve onto the disk, we
can find the equation of an arbitrary geodesic (on the disk) that does
not include the origin:

Because the boosted $x_3'$ and $x_3''$ axes define the plane whose
intersection with the paraboloid determines the new geodesic, this plane
must lie within the light cone. It follows that there is a normal to the
plane making an angle $\theta$ with the original $x_3$ axis, with
$\pi/2>\theta>\pi/4$:
\begin{equation}
0<\cos\theta<\cos\frac{\pi}{4}=\dfrac{1}{\sqrt{2}}. \label{E22g}
\end{equation}
On the other hand, if we express the normal to the plane
as $(a,b,1)$, then any point $(x_1,x_2,x_3)$ on the plane satisfies
\begin{equation}
(a,b,1)\cdot(x_1,x_2,x_3)=0, \label{E22h}
\end{equation}
or $ax_1+bx_2+x_3=0$.
However,
Eq.~(\ref{E22g}) tells us that
\begin{equation}
\cos\theta
=\dfrac{(a,b,1)\cdot(0,0,1)}{\sqrt{a^2+b^2+1}}<\dfrac{1}{\sqrt{2}},
\end{equation}
which implies that $\sqrt{a^2+b^2+1}>\sqrt{2}$ and $a^2+b^2>1$. We thus
conclude that the plane shown in Fig. is specified by
\begin{equation}
ax_1+bx_2+x_3=0 \quad \text{with} \quad
a^2+b^2>1.\label{E23}
\end{equation}
Because any such plane can be obtained by boosting the appropriate
$x_3'$ and $x_3''$, we conclude that planes of the form in
Eq.~(\ref{E23}) determine all the geodesics on the paraboloid. That is,
a curve is a geodesic if and only if it lies on the intersection of the
paraboloid and any plane including the origin.

Our main interest is in the points shown in Fig. that lie
on the intersection of the plane and the paraboloid. Any point on the
paraboloid satisfies Eq.~(\ref{E29k}),
\begin{equation}
x_3=\dfrac{1+x_1^2+x_2^2}{2}.
\label{87}
\end{equation}
If we combine Eq.~(\ref{87}) with Eq.~(\ref{E23}), we see that points
on the curve formed at the intersection of the plane and the paraboloid
(that is, points on a geodesic on the paraboloid that do not pass
through its vertex) satisfy
\begin{align}
&-ax_1-bx_2=\dfrac{x_1^2+x_2^2+1}{2}, \qquad (a^2+b^2>1)\\
\noalign{\noindent which implies that}
& (x_1+a)^2+(x_2+b)^2 = a^2+b^2-1>0.\label{E23p}
\end{align}

If we project back down onto the disk as shown in Fig., we
see that any geodesic that does not include the origin is the arc of a
circle centered at $(-a,-b)$ with radius $\sqrt{a^2+b^2-1}$. Note that
the center of any of these circles always lies outside of the unit disk
because
$a^2+b^2>1$, and that any point $(a,b)$ outside of the unit disk is the
center of a circle on which some geodesic inside the disk lies. Also,
because $a$ and $b$ can be positive or negative, it doesn't really matter
whether we denote the center of the circle by $(-a,-b)$ or
$(a,b)$.

Finally, we can prove that the circles derived in Eq.~(\ref{E23p}) are perpendicular to the edge of the disk at their
points of intersection. We recall from Sec.~\ref{S5} that points
on the disk are denoted by
$(x,y)$, points on the edge of the disk satisfy
\begin{equation}
x^2+y^2=1,\label{E24}
\end{equation}
and points on a
geodesic on the disk satisfy
\begin{equation}
(x+a)^2+(y+b)^2 =
a^2+b^2-1.\label{E25}
\end{equation}
By taking the differential of 
Eq.~(\ref{E24}) and Eq.~(\ref{E25}), we obtain
\begin{align}
2xdx+2ydy&=0 \quad \label{E26} \\ 2(x+a)dx+&2(y+b)dy=0,
\end{align}
which can be rewritten as
\begin{align}
\dfrac{dy}{dx}&=-\dfrac{x}{y} \label{E27} \\
\dfrac{dy}{dx}&=-\dfrac{x+a}{y+b}.\label{E28}
\end{align}
Equation~(\ref{E27}) gives the slope of the tangent to any point on the
edge of the disk, and Eq.~(\ref{E28}) gives the slope of the tangent to
any point on the geodesic. To prove that the geodesic is perpendicular
to the edge of the disk at their point of intersection, we must show
that these two tangents are perpendicular to each other; that is, that
\begin{equation}
-\dfrac{x}{y}=\dfrac{y+b}{x+a}. \label{E29}
\end{equation}
To do this, we note that any point on both the edge of the
disk and on the geodesic satisfies both Eqs.~({\ref{E23p}) and
(\ref{E24}), and by adding these two equations together we find
\begin{align}
2x^2+2ax+2y^2+2by &= 0 \\ 
x(x+a)+y(y+b)&=0 \\
-\dfrac{x}{y}=\dfrac{y+b}{x+a}&,
\end{align}
as required. Note that this argument can be reversed to show that any
circular arc orthogonal to the unit circle at their points of
intersection is a geodesic.

%\begin{figure}
%\begin{center}
%\includegraphics[width=3.5in, height=2in]{rhodes6.eps}
%\end{center}
%\caption{Three geodesics in rapidity space and a non-Euclidean triangle.  The geodesics are labelled as $\Phi_i$ and the angles of intersection as $\alpha_i$}
%\label{F6}
%\end{figure}

The main results of this section are the following. As shown in
Figs. and, geodesics through the origin of rapidity space are straight
lines (in the Euclidean sense), and any straight line through the
rapidity space origin is a geodesic.

As shown in Figs. and , geodesics in rapidity
space which do not include the origin are arcs of circles whose centers
lie outside the disk and are perpendicular to the edge of the disk
at their points of intersection.  Conversely, any
point outside the disk is the center of some circular arc within the
disk that is a geodesic. As Fig. shows, the new feature is that
rapidity space contains geodesics that are not ``straight'' in
the Euclidean sense. One way to understand these curved geodesics is to
note that the metric of Eq.~(\ref{allequations28c}) tells us that
segments near the edge of the disk are much longer then they appear, so
the shortest path between two points near the disks' edge must be bowed
inward rather than straight.

\section{Applications}\label{S7}
Now that we have found the Lorentz invariant
metric and geodesics in rapidity space, we can investigate the
relativistic addition of velocities and the various
consequences of successive non-colinear Lorentz boosts.

\subsection{Qualitative Results}\label{S7a}

Before deriving quantitative expressions for the Thomas-Wigner rotation
and Thomas precession, we first discuss results that can be
deduced geometrically without the use of any equations.

Velocities are represented by points in rapidity space. A boost from
one velocity to another is represented in rapidity space by the geodesic
connecting them. Because a pure boost doesn't involve any rotation of the
reference frame being boosted, the coordinate axes representing the boost
in rapidity space maintain a fixed angle with respect to the geodesic
they follow, as shown in Fig..  When coordinate axes maintain a fixed angle with a geodesic as they move along it from one point to another, they are said to undergo \emph{parallel transport}.  The most familiar example of parallel transport occurs in nonrelativistic velocity space, in which all the geodesics are straight lines.

We begin by considering a set of colinear boosts in real space. If we
assume this set contains the zero velocity frame, then the first boost
is represented in rapidity space by a segment of a straight line
geodesic that includes the rapidity space origin. Without any loss of
generality, we take that direction as the horizontal axis in both real
and rapidity space. Because straight lines through the origin of rapidity
space are geodesics, each colinear boost is represented by a segment of
the same (horizonal) line. If we represent a coordinate system in
rapidity space by two small perpendicular lines (crosshairs) centered on
the point of interest, then as shown in Fig., when we boost
from one velocity to another, the orientation of the crosshairs remains
fixed with respect to the geodesic connecting them. Thus, the orientation of the crosshairs is unchanged no matter how many colinear boosts it undergoes because, in each case, it
is moving along the same straight line geodesic in rapidity space.

%\begin{figure}
%\begin{center}
%\includegraphics[width=3.5in, height=1.5in]{rhodes7.eps}
%\end{center}
%\caption{The first boost, along the geodesic $\Phi_1$.}\label{F7}
%\end{figure}

Furthermore, because the distance (as measured with the rapidity space
metric) from the origin to any point in rapidity space is the rapidity
of that point, we see that when successive boosts are colinear, the
corresponding rapidities add and subtract like ordinary numbers. Thus,
rapidity space provides an easy geometrical way to obtain
Eq.~(\ref{allequations6e}) and to prove that frames boosted in the same
direction do not rotate with respect to each other. Working in rapidity
space also provides an easy proof that no matter how many
colinear velocities are added together, the magnitude of their sum
always will be less than the speed of light.

A boost that does not include the zero velocity frame corresponds to
a geodesic in rapidity space that does not include the rapidity space
origin. As we proved in Sec.~\ref{S6}, this geodesic lies on the
arc of a circle whose center is outside the disk. As shown in
Fig., crosshairs moving along this type of geodesic maintain
their orientation with respect to it. Therefore, crosshairs moving back
to the origin along a closed path that includes one (or more) of these
geodesics will be rotated with respect to their initial
orientation. An example of this rotation is shown in
Fig.. Suppose we boost from rest to a velocity $\vec{v}$
along the
$x$-axis (in real space). Then we perform a non-colinear boost from a
frame with velocity
$\vec{v}$ to a frame with velocity $\vec{v'}$, and finally, we boost
from a frame with velocity $\vec{v'}$ back to the original rest frame.
If we look at the corresponding points in rapidity space as shown in
Fig., we see that the frame obtained at the end of these three
boosts is rotated with respect to the one that stayed at the origin.
This rotation is the Thomas-Wigner rotation, which we denote
by TWR, and the geometry of rapidity space shows that the TWR is in the
clockwise (negative) direction when a frame is moving in
rapidity space in the counterclockwise (positive) direction (and
vice versa).

%\begin{figure}
%\begin{center}
%\includegraphics[width=3.5in, height=1.5in]{rhodes8.eps}
%\end{center}
%\caption{The second boost, along the geodesic $\Phi_2$.}\label{F8}
%\end{figure}

%\begin{figure}
%\begin{center}
%\includegraphics[width=3.5in, height=1.5in]{rhodes9.eps}
%\end{center}
%\caption{The third boost, along the geodesic $\Phi_3$.}\label{F9} 
%\end{figure}

It also is easy to see that there is an upper limit on the TWR
angle. Without any loss of generality, suppose we first boost along the
$x$-axis (in real space) to a frame whose speed is very close to the
speed of light. There is no orientation change of the boosted frame
because the geodesic it follows in rapidity space is a straight line. If
we next perform a non-colinear boost to a speed even closer to the speed
of light, which makes an angle slightly less than $\pi$ with the
$x$-axis, then as shown in Fig., the geodesic
representing this second boost will lie on the arc of a circle that is
perpendicular to the edge of the disk (where $v=c$) at its two points of
intersection. In the limit that both speeds approach the speed of light
and the angle between the boosts approaches $\pi$, the arc representing
the second boost approaches a half circle. Thus, the change in orientation of a reference frame following the arc approaches $\pi$. Consequently, any Thomas-Wigner rotation angle has an upper limit of $\pi$, and the limit of $\pi$ is approached only when the two boosts involved have speeds very close to the speed of light and are almost opposite to each other. In
all other cases the TWR angle will be less than $\pi$.

%\begin{figure}
%\begin{center}
%\includegraphics[width=3.5in, height=2.0in]{rhodes10.eps}
%\end{center}
%\caption{When a
%second boost is applied at an angle approaching $\pi$ with the first, the
%Thomas-Wigner rotation angle approaches its maximum value of
%$\pi$.}\label{figmaxTP.eps} 
%\end{figure}

On the other hand, if we perform the same two non-colinear boosts as
before, but now give each a nonrelativistic speed ($\beta\ll 1$), then
as shown in Fig., even though the geodesic representing the
second boost still lies on the arc of a circle, that arc is indistinguishable from a straight line because it is located near the rapidity space origin. Therefore, when the boost speeds are nonrelativistic, there is essentially no rotation of the frame in following the two geodesics.

Next consider a reference frame in real space undergoing circular
motion in the counterclockwise (positive) direction with a constant,
nonrelativistic speed. Classically, this situation is treated by
representing it as the limiting case of a set of small, non-colinear
boosts. That is, circular motion is approximated as motion along a
polygon with an ever increasing number of sides. Because the boosts
involved all have the same nonrelativistic speed, all the
circular arcs representing them in rapidity space are essentially
indistinguishable from straight lines. Hence, as long as the speed of
the frame in circular motion is nonrelativistic, it undergoes essentially
no change in orientation upon its return to the origin (as we would
expect).

%\begin{figure}
%\begin{center}
%\includegraphics[width=3.5in, height=2in]{rhodes11.eps}
%\end{center}
%\caption{All of these
%non-Euclidean triangles have the same two base angles, but the smaller a
%triangle is, the closer it is to appearing Euclidean.}\label{F10}
%\end{figure}

Now suppose the reference frame undergoing circular motion has a
constant speed that is relativistic. The geodesics in rapidity
space that form the polygon are now small
arcs that lie on circles whose intersection with the edge of the disk
is orthogonal, as shown in Fig.. Consequently, the frame being boosted undergoes a definite change in orientation with each boost. Thus, when the reference frame returns to its starting point, it will have undergone a clockwise (negative) rotation with 
respect to its initial orientation. This sum of all the rotations experienced along the way is the Thomas
precession. Furthermore, we see from the geometry of rapidity space
that the amount of rotation will be a function of the speed of the
circular motion (that is, the rapidity space distance from the
rapidity space origin), and will increase without bound as this speed
approaches the speed of light, as shown in Fig..

%\begin{figure}
%\begin{center}
%\includegraphics[width=3.5in, height=2in]{rhodes12.eps}
%\end{center}
%\caption{Polygonal
%approximations to curved paths in rapidity space.}\label{F11}
%\end{figure}

We also can use the geometry of rapidity space to show that non-colinear, 
relativistic boosts are also (in general) noncommutative. Suppose we first boost a reference frame from speed zero to a speed close to the speed of light, and then boost the frame through
a second non-colinear velocity. From the rapidity space diagram in
Fig., it is clear that we will end up at a completely
different point if we do the same boosts in reverse order. However, if
we look closely at the rapidity space diagram, we see that although the
two resultant velocities are represented by different points in rapidity
space, they both are the same (rapidity space) distance from the
rapidity space origin. Consequently, they both have the
same rapidity (and hence speed), but not the same direction. The result
that the final speed resulting from two successive non-colinear boosts
is independent of the order in which the boosts are applied is
usually proved by a somewhat long algebraic calculation.\cite{ftnote2}

%\begin{figure}
%\begin{center}
%\includegraphics[width=3.5in, height=2in]{rhodes13.eps}
%\end{center}
%\caption{Two boosts, with corresponding rapidities $\phi_i$
%in non-parallel directions, do not commute.  In both cases, the angle between the boosts is $\alpha$.}\label{F12} \end{figure}

Finally, it is easy to see that the sum of any number of non-colinear
relativistic velocities always results in a velocity whose magnitude is
less than the speed of light. Although obvious when viewed on a
rapidity space diagram, the corresponding algebraic proof is somewhat
complex.

\subsection{Relation of the Thomas-Wigner rotation to the rapidity
space triangle}

Because angles in rapidity space behave exactly like angles in Euclidean
space, it is relatively easy to quantify the arguments of
Sec.~\ref{S7a}. As shown
in Fig., the angles between the boosts are
$\alpha_3$, $\alpha_1$, and $\alpha_2$. The straight line
geodesic making an angle $\alpha_3$ with the horizontal axis is
called $\Phi_3$, and the other two geodesics are $\Phi_1$ and
$\Phi_2$. The length (rapidity) of the segment of the geodesic
$\Phi_i$ representing each boost is denoted by $\phi_i$. We see
in Fig. that when we boost from the origin along $\Phi_1$,
there is no change in the orientation of the crosshairs. As shown in
Fig., when we boost along $\Phi_2$, the $x$-axis of the
crosshairs maintains its angle $\pi-\alpha_1$ with respect to $\Phi_2$
because $\Phi_2$ is a geodesic. Finally, as shown in Fig., boosting
back to the origin along $\Phi_3$, the $x$-axis of the crosshairs
maintains its orientation of \;$(\pi-\alpha_1)-\alpha_2$ with respect to
the geodesic $\Phi_3$. Thus, as Fig. shows, when the
coordinate system returns to the rapidity space origin, its $x$-axis
will have rotated from its initial orientation in the clockwise
(negative) direction by the Thomas-Wigner rotation angle
\begin{equation}
{\rm TWR} = -\big[\pi -(\alpha_1 +\alpha_2
+\alpha_3)\big].
\label{E32a}
\end{equation}

Note that if the two rapidities $\phi_1$ and $\phi_2$ are small (that is, if
the speeds they represent are nonrelativistic) then, as shown in Fig., the
figure formed by the segments of the three geodesics $\Phi_i$ is indistinguishable from a Euclidean triangle (because in this
case $\alpha_1 + \alpha_2+ \alpha_3\approx \pi)$ and, as expected, the
TWR angle is essentially zero. On the other hand, if the two rapidities $\phi_1$ and $\phi_2$
are large, then the resulting TWR angle can approach the upper limit of
$\pi$, as discussed above.

The absolute value of the right-hand side of Eq.~(\ref{E32a}) is known
to mathematicians as the ``angular
defect,'' because it is a measure of how much the sum of
the angles inside a triangle differs from the corresponding sum in
ordinary Euclidean space ($\pi$). The theorem that the angular defect of
a triangle is equal to its area is proven in hyperbolic geometry
courses. Rather than simply invoking this result, we can establish it from first
principles. Even if we had no previous knowledge of hyperbolic
geometry, we might suspect that the area enclosed by the rapidity space
triangle is proportional to the TWR angle, because our method for
finding the TWR angle involves traveling around a closed path and
summing up the angular change along the way. This sum corresponds to evaluating an integral of the form
\begin{equation}
\oint\limits_C d\theta. \label{E32}
\end{equation}
An integral like (\ref{E32}) appears in Green's theorem, which relates an
integral around a closed curve to an integral over the two-dimensional
area enclosed by that curve. We now show that the integral
Eq.~(\ref{E32}) is actually the left-hand side of Green's theorem for a
particular choice of the integrand, and that this choice makes the
right-hand side of Green's theorem equal to the area enclosed by the
curve.

We begin by writing Green's theorem as
\begin{equation}
\oint\limits_C
\vec{F} \cdot d\vec{s}=\iint\limits_\Sigma\nabla\times
\vec{F}\cdot\hat{n}\,d\sigma,\label{E33}
\end{equation}
where $C$ signifies any closed (2D) curve traversed in the
counterclockwise direction, $\hat{n}$ is the unit vector normal to the
plane of this curve (according to the right-hand rule), $\Sigma$
stands for the region enclosed by the curve, and $\vec{F}$ is any
vector defined on rapidity space.

To apply Green's theorem in rapidity space, we first must find explicit
expressions for each integrand. To do this, recall that
\begin{align}
ds^2 &=\dfrac{4}{(1-r^2)^2}\Bigl(dx^2+dy^2\Bigr)=h_1^2\; dx^2 + h_2^2\;
dy^2, \\
\noalign{\noindent and}
d\vec{s}&=\Bigl(\dfrac{2}{1-r^2}\Bigr)dx\:\hat{\imath}+
\Bigl(\dfrac{2}{1-r^2}\Bigr)dy\,\hat{\jmath}.\label{E33a} \end{align}
From Eq.~(\ref{E33a}), we find that the area
element in Cartesian and plane polar coordinates is
\begin{align} d\sigma
&=\Bigl[\Bigl(\dfrac{2}{1-r^2}\Bigr)dx\Bigr]\Bigl[\Bigl(\dfrac{2}{1-r^2}\Bigr)dy\Bigr]
\\ &=\Bigl(\dfrac{2}{1-r^2}\Bigr)^2dx\,dy =
\Bigl(\dfrac{2}{1-r^2}\Bigr)^2\,rdrd\theta. \label{I2} \end{align}

We also can give a more informal derivation of Eq.~(\ref{I2}). Because
the metric Eq.~(\ref{E33a}) is conformal, it is locally a multiple of
the Euclidean metric (even though this multiple varies from point to
point).
 Therefore, we expect the area element also to be a multiple of the
Euclidean area element, with the multiplying factor equal to
$\big(2/{(1-r^2)}\big)^2$, because the area is the product of the
infinitesimal length in each of the two orthogonal directions and each
length is the Euclidean length multiplied by the factor $2/(1-r^2)$.

To evaluate the integral on the right-hand side of Eq.~(\ref{E33}), we
need to express the curl and dot product in the coordinates of rapidity
space. From Boas\cite{Boas} we have
\begin{subequations}
\begin{eqnarray}
\nabla\times \vec{F}\cdot\hat{n}&=& \dfrac{1}{h_1
h_2}\Bigl[\dfrac{\partial}{\partial x}\Bigl(h_2 F_2\Bigr) -
\dfrac{\partial}{\partial y}\Bigl(h_1 F_1\Bigr)\Bigr] \\
&=&\Bigl(\dfrac{1-r^2}{2}\Bigr)^2\dfrac{\partial}{\partial
x}\Bigl[\Bigl(\dfrac{2}{1-r^2}\Bigr)F_y\Bigr] -
\dfrac{\partial}{\partial
y}\Bigl[\Bigl(\dfrac{2}{1-r^2}\Bigr)F_x\Bigr]. \label{E34}
\end{eqnarray}
\end{subequations}
Because Green's theorem is true for any vector $\vec{F}$, it holds
for the particular vector
$\vec{F}=-y \hat{\imath}+x \hat{\jmath}$. For this choice,
the right-hand side of Eq.~(\ref{E34}) reduces to $1$.

Thus, when $\vec{F}=-y \hat{\imath}+x \hat{\jmath}$, Green's theorem
becomes
\begin{equation}
\mbox{Area} (\Sigma) = \oint\limits_C \vec{F} \cdot
d\vec{s}, \label{E35}
\end{equation}
where the area enclosed by the
closed curve $C$ is calculated using the rapidity space area element
Eq.~(\ref{I2}).  If we substitute our choice for $\vec{F}$ into the right-hand side of
Eq.~(\ref{E35}), we find
\begin{subequations}
\begin{eqnarray}
\oint\limits_C \vec{F} \cdot
d\vec{s} &=&\oint\limits_C (-y\hat\imath + x\hat\jmath)\cdot
\Bigl(\dfrac{2}{1-r^2}\Bigr)\Bigl(dx\hat\imath +dy\hat\jmath\Bigr) \\
&=&\oint\limits_C\Bigl(\dfrac{2}{1-r^2}\Bigr)\Bigl(xdy-ydx\Bigr).
\label{E36}
\end{eqnarray}
\end{subequations}
Therefore, Eq.~(\ref{E35}) can be written as
\begin{equation}
\mbox{Area} (\Sigma) = \oint\limits_C
\Bigl(\dfrac{2}{1-r^2}\Bigr)(x\,dy-y\,dx).\label{E37}
\end{equation}

We now show that the right-hand side of Eq.~(\ref{E37}) is equal to
an integral of the form given in Eq.~(\ref{E32}). Let $r$ and $\theta$
be the usual plane polar coordinates measured from the center of the
disk. In terms of these coordinates,
\begin{subequations}
\begin{align}
x&=r \cos\theta \quad \text{and} \quad dx = -r\sin\theta d\theta+dr
\cos\theta, \\ y&=r\sin\theta \quad \text {and} \quad dy = \quad
r\cos\theta d\theta +dr \sin\theta.
\end{align}
\end{subequations}
We substitute these coordinates
into Eq.~(\ref{E37}) and find that
\begin{align}
 &x\,dy-ydx = r^2 d\theta, \\
\noalign{\noindent which implies that}
& \mbox{Area}(\Sigma) =
\oint\limits_C\left(\dfrac{2r^2}{1-r^2}\right) d\theta.
\label{E38}
\end{align}

By looking at the hyperbolic triangle in Fig. representing the
three boosts, we see that two of the three sides are straight lines
emanating from the origin, which means that $d\theta=0$ for these
geodesic segments.  Consequently, the integral along each of these segments makes no
contribution to the path integral. Therefore, only the integral over the
curved geodesic contributes to the right-hand side of Eq.~(\ref{E38}). We can evaluate this integral by changing it from one in terms of the polar
coordinates
$(r,\theta)$ measured from the center of the disk to an integral in
terms of coordinates measured from the center
$(a,b)$ of the circle on which the curved arc lies (see
Fig.). If
$\sqrt{a^2 + b^2 -1}$ is the radius of this circle and $\omega$ the
corresponding angular coordinate (defined as positive in the
counterclockwise direction), then the desired coordinate transformation
is
\begin{subequations}
\begin{align}
x &= a - \sqrt{a^2 + b^2 -1} \cos\omega, \\ y &= b -
\sqrt{a^2 + b^2 -1} \sin\omega.
\end{align}
\end{subequations}
After some algebra, we find that
\begin{equation}
\dfrac{2}{1-r^2}=
\dfrac{1}{1-(a^2 + b^2)+\sqrt{a^2 + b^2 -1}\big(a\cos\omega +
b\sin\omega\big)},\label{E39}
\end{equation}
and
\begin{equation}
x dy-y dx =-\big[1-(a^2 +
b^2)+\sqrt{a^2 + b^2 -1}\big(a\cos\omega +
b\sin\omega\big)\big]d\omega.\label{E40} 
\end{equation}
If we use Eqs.~(\ref{E39}) and (\ref{E40}) in the integrand of
Eq.~(\ref{E37}), we conclude that
\begin{equation}
\Bigl(\dfrac{2}{1-r^2}\Bigr)(x dy-y dx)= -d\omega.\label{E41a}
\end{equation}
Using Eq.~(\ref{E41a}) in the integral in Eq.~(\ref{E37}), we see that the integral in Eq.~(\ref{E37}) is zero on the straight lines through the origin, while on the other geodesic, it is (to within a sign) the angular extent of the geodesic segment about its center.

%\begin{figure}
%\begin{center}
%\includegraphics[width=3.5in, height=2in]{rhodes14.eps}
%\end{center}
%\caption{The
%non-Euclidean area of a non-Euclidean triangle is equal to its angular
%defect $\pi-(\alpha_1 +\alpha_2 +\alpha_3) = \omega$.}
%\label{F13}
%\end{figure}

If we look at Fig., we see that the two radii of the circle
centered at $(a,b)$ together with the two sides of the triangle which are
straight lines, form a four-sided figure (that is, a Euclidean
quadrilateral). Because the sum of the angles in a Euclidean quadrilateral
is $2\pi$,\cite{ftnote15} we have that
\begin{subequations}
\begin{align}
2\pi &= \alpha_3 + \Bigl(\alpha_1 + \dfrac{\pi}{2}\Bigr) +
\omega + \Bigl(\alpha_2 + \dfrac{\pi}{2}\Bigr) \\ &=(\alpha_1 +
\alpha_2 + \alpha_3)+ \omega + \pi,
\end{align}
\end{subequations}
which implies that
\begin{equation}
\omega = \pi - (\alpha_1
+ \alpha_2 + \alpha_3).
\end{equation}
Therefore, the area enclosed by the
triangle is
\begin{subequations}
\begin{align}
\mbox{Area} (\Sigma) =
-\int_{\omega_1}^{\omega_2}d\omega&= \int_{\omega_2}^{\omega_1} d\omega
\\ =\pi - (\alpha_1& + \alpha_2 + \alpha_3) =-({\rm TWR}).\label{E41}
\end{align}
\end{subequations}
Note that because we have proved the area of our special triangle
is just the angular sweep of its one curved side, it is easy to prove that the area of any geodesic-sided polygon is the sum of the angular sweeps of its
sides (about their various centers of curvature).

We have thus proved the main result of this section, that the
negative of the Thomas-Wigner rotation is equal to both the
(rapidity space) area enclosed by the rapidity space triangle and the
angular defect ($\pi$ minus the sum of the interior angles of the
rapidity space triangle). Although we have derived this result from
first principles, it was pointed out by Aravind,\cite{Aravind} and later
discussed in a slightly different context by Criado and
Alamo.\cite{Criado}

Although the result (\ref{E41}) is interesting in its own right, it also suggests
that another way to evaluate and compare Thomas-Wigner rotations is to
look at the areas of the corresponding triangles in rapidity space. 
Although possible in principle, in practice this is not very easy to do
because areas in rapidity space depend on where they are located, and
thus are not readily compared using our Euclidean-trained eyes. More
specifically, as Eq.~(\ref{allequations28c}) for the area element
$d\sigma$ shows, although two regions in different parts of rapidity
space may appear to have the same area to our Euclidean-trained eyes,
the area of the one closest to the edge of the disk is larger.

\subsection{Various Equations for the Thomas-Wigner Rotation Angle}

We now are in a position to derive the various expressions for the
Thomas-Wigner rotation angle that have appeared in the
literature. Because most of these expressions are for the magnitude of the
TWR, we can use Eqs.~(\ref{E37}) and (\ref{E41}) to write
\begin{subequations}
\begin{align}
\Bigl|\mbox{TWR}\Bigr| & =\Bigl|
\oint\Bigl(\dfrac{2}{1-r^2}\Bigr)\Bigl(x dy - y dx\Bigr)\Bigr| \\
&=\oint\Bigl(\dfrac{2}{1-r^2}\Bigr)|(\vec{r}\times d\vec{r})\cdot\hat k|
\\
&=\oint\Bigl(\dfrac{2r^2}{1-r^2}\Bigr)\Bigl|\dfrac{(\vec{r}\times
d\vec{r})\cdot\hat k}{r^2}\Bigr|.\label{E42}
\end{align}
\end{subequations}
If we use
Eqs.~(\ref{E21b}) and (\ref{I1}), we find that
\begin{equation}
\dfrac{2r^2}{1-r^2}=\gamma - 1.\label{E42e} \end{equation}
Equation~(\ref{E21h}) can then be used to show that for any infinitesimal
segment of the path,
\begin{equation}
\left|\dfrac{(\vec{r}\times
d\vec{r})\cdot\hat k}{r^2}\right|= \left|\dfrac{(\vec{v}\times
d\vec{v})\cdot\hat k}{v^2}\right|,
\end{equation}
which means that
Eq.~(\ref{E42}) can be rewritten as
\begin{equation}
\left|\mbox{TWR}\right|=\oint_C (\gamma -1)\left|\dfrac{\vec{v}\times
d\vec{v}}{v^2}\right|.\label{E43b}
\end{equation}

Equation~(\ref{E43b}) can be re-expressed in various forms. For example,
if we call the integrand $d\chi$, then
\begin{subequations}
\begin{eqnarray}
d\chi &=&
\dfrac{\gamma -1}{v^2}\left|\vec{v}\times d\vec{v}\right| \label{E43}\\
&=&\dfrac{\gamma -1}{v^2}\left|\vec{v}\times \vec{a}\right|dt
\end{eqnarray}
\end{subequations}
Hence,
\begin{equation}
\dfrac{d\chi}{dt}=\dfrac{\gamma
-1}{v^2}\left|\vec{v}\times
\vec{a}\right|.\label{E43a}
\end{equation}
Equation~(\ref{E43}) is the
expression for the Thomas-Wigner rotation angle given in
Ref.~\onlinecite{Sard}, p.~289 and Ref.~\onlinecite{Arzelies}, p.~178.

Several interesting physical properties can be deduced from
Eq.~(\ref{E43b}). First, the right-hand side tends to zero in the
nonrelativistic limit, showing that in this limit the Thomas-Wigner
rotation vanishes. Second, as $v\bigarrow c$, the
Thomas-Wigner rotation angle increases without bound, as we 
deduced in Sec.~\ref{S7a} using the geometry of rapidity space. Third,
the Thomas-Wigner rotation is a purely kinematic effect because it is
independent of the dynamics causing the acceleration. In other words, it
not only occurs for charged particles moving in electromagnetic fields,
but also can occur for elementary particles accelerated by nuclear
forces,\cite{Jackson3} and for masses accelerated by gravitational
fields.

If we multiply the right-hand side of Eq.~(\ref{E43a}) by
$(\gamma+1)/(\gamma+1)$ and use the identity given in Eq.~(\ref{I1}), we
find that
\begin{equation}
\label{E44} 
\omega=
\dfrac{d\chi}{dt}=\dfrac{\gamma^2}{\gamma+1}\Bigl|
\vec{\beta}\times\dfrac{d\vec{\beta}}{dt}\Bigr|
=\dfrac{(\gamma-1)}{\beta^2}\Bigl|\vec{\beta}
\times\dfrac{d\vec{\beta}}{dt}\Bigr|,
\end{equation}
which is the expression for the angular speed of the
Thomas-Wigner rotation given in Ref.~\onlinecite{Sard},
p.~290, Ref.~\onlinecite{Munoz}, p.~554, and
Ref.~\onlinecite{Arzelies}, p.~179. Because it is
the angular velocity of the Thomas-Wigner rotation that enters into
the calculation of the Thomas precession, it is this quantity that
appears in the relativistic correction to the spin-orbit term in the
Hamiltonian for a hydrogen atom.

\subsection{Applying the Thomas Precession in Quantum Theory}

Most derivations of the relativistic correction to the spin-orbit term
in the Hamiltonian for a hydrogen atom relate the time rate of change
of the electron's spin vector in its instantaneous rest frame to the
corresponding rate in the lab (or proton's rest) frame.\cite{Munoz}
Because any instantaneous rest frame of the electron is obtained from
the previous instantaneous rest frame by a non-colinear Lorentz boost, the
transformation back to the lab frame will include Thomas-Wigner
rotations. The rate at which the Thomas-Wigner rotations occur is the
rate given in Eq.~(\ref{E43a}). There are several excellent
derivations of the correct form of the relativistic correction to the spin-orbit term (see, for
example, Refs.~\onlinecite{Fisher}, \onlinecite{Jackson}, or
Ref.~\onlinecite{Griffiths}), and Eq.~(\ref{E43a}) is used in all of them.

Some authors\cite{Haken} claim the factor of two that comes
from including the Thomas precession in the spin-orbit term results
from the electron's rest frame precessing through one complete cycle
each time it completes one revolution around the proton. However, as
Eqs.~(\ref{E44}) and (\ref{E44f}) show, this interpretation is
incorrect because the number of rotations completed during each
revolution is variable, and tends to infinity as
$v\bigarrow c$.

\subsection{A Special Case of Thomas Precession}

The Thomas precession of an object is the sum of all the Thomas-Wigner
rotations it undergoes when it completes one closed planar orbit. To see
this explicitly, consider the simple example of an object moving in a
circle with a constant speed. If we use the area element given in
Eq.~(\ref{I2}), the expression for the magnitude of the Thomas
precession for this case is
\begin{subequations}
\begin{align} \big|{\rm TP}\big|
&=\int_0^{2\pi}\int_0^R\dfrac{4}{(1-r^2)^2}r dr d\theta \\ &=
2\pi\int_0^R\dfrac{4r}{(1-r^2)^2}dr.
\end{align}
\end{subequations}
We can evaluate this
integral by changing the integration variable to $u=(1-r^2)$. After some
algebra, we find
\begin{equation}
\big|{\rm TP}\big| =
4\pi\Bigl(\dfrac{R^2}{1-R^2}\Bigr).
\end{equation}
By using
Eq.~(\ref{E42e}), we see that
\begin{equation}
\Bigl(\dfrac{R^2}{1-R^2}\Bigr)=\dfrac{\gamma - 1}{2}, \end{equation}
which means that
\begin{equation}
\big|{\rm TP}\big|=2\pi(\gamma -
1)\label{E44f}.
\end{equation}
Note that if the object moves around the circle in the clockwise
(negative) direction, then the Thomas precession is in the opposite
(positive) direction after one revolution around the circular path.
This result also is derived in Ref.~\onlinecite{Arzelies}, p.~179.

\section{Mathematical Connections and Alternative Equations for the
Thomas-Wigner Rotation}\label{S8}

The purpose of this section is to give a brief discussion
of the relation between the results presented in this paper and
M\"obius transformations, spinors, the group $SL_2(\mathbb C)$, and
models of the hyperbolic plane. The only new physical result is
Eq.~(\ref{A4}), which expresses the Thomas-Wigner rotation angle in
terms of the rapidities that give rise to it and the angle between their
corresponding boosts. Equation~(\ref{A4}) also can be obtained
geometrically in rapidity space; it is just easier to derive in the
present context.

We have shown that rapidity space and the actions of Lorentz
transformations on it provide valuable insight into the
Thomas-Wigner rotation and the Thomas precession. Although our
presentation has not required an actual algebraic expression for
the action of Lorentz transformations on rapidity space, it is natural
to ask for one. Once we have this expression, it will be easy to relate
Lorentz transformations to certain M\"obius transformations (linear
fractional transformations), and then to the spinor map between
$SL_2(\mathbb C)$ and the Lorentz group.

\subsection{Lorentz Transformations of Rapidity Space}\label{S8a}

To see how the Lorentz transformation of Eq.(\ref{E6}) (a boost in the
positive $x$-direction with rapidity $\phi$) acts on a point $(x,y)$ in
the Poincar\'e disk, let $(x,\,y,\,(x^2+y^2+1)/2)$ denote the point on
the paraboloid that projects to this point in the disk (see
Fig.). If we apply the boost to this vector, we obtain
\begin{equation}
\begin{pmatrix} x \\ y\\ \frac{x^2+y^2+1}{2}
\end{pmatrix}
\mapsto
\begin{pmatrix}
(\cosh \phi )x -\sinh
\phi\big(\frac{x^2+y^2+1}2\big)\\ y\\ -(\sinh \phi)x +\cosh \phi
\big(\frac{x^2+y^2+1}2\big)
\end{pmatrix},
\end{equation}
which we
then need to rescale so that it lies on the paraboloid. Some rather
messy algebra shows the correct scaling factor is 
\begin{equation}
\lambda=\Bigl( \frac
{\cosh \phi +1}2 -(\sinh \phi) x +\frac {\cosh \phi-1}2 (x^2 + y^2)
\Bigr)^{-1},
\end{equation}
and thus the boost maps points in the Poincar\'e
disk by
\begin{equation}
\begin{pmatrix} x\\ y
\end{pmatrix}
\mapsto
\begin{pmatrix}
x'\\ y' \end{pmatrix}= \begin{pmatrix} \lambda
\big((\cosh \phi) x -\sinh \phi (\frac {x^2+y^2+1}2)\big)\\
\lambda y
\end{pmatrix}.
\label{132}
\end{equation}
Equation~(\ref{132}) can be expressed in a
surprisingly simple way if we use complex notation to denote points in
the disk. If we let $z=x+iy$, and set
\begin{eqnarray}
a&=&\sqrt{\frac {\cosh \phi+1}2}=\cosh \frac \phi 2\label{A10}\\
b&=&-\sqrt{\frac {\cosh \phi-1}2}=-\sinh \frac \phi
2,\label{A11}
\end{eqnarray}
the action of the boost becomes
\begin{equation}
z\mapsto
z'=\frac {az+b}{bz+a}.
\end{equation}
Thus, when we use complex notation to label
points on the Poincar\'e disk, the action of the boost can be expressed
in a particularly simple way as a M\"obius transformation.

Another special type of Lorentz transformation that is easy to analyze
is a spatial rotation. It is not difficult to see that a
counterclockwise spatial rotation by an angle of $\theta$ produces the
map of the disk
\begin{equation}
z\mapsto z'=e^{i\theta}z= \frac
{e^{i\theta/2}z+0}{0 z+e^{-i\theta/2}},
\end{equation}
which is again a M\"obius transformation.

Given any M\"obius transformation $z\mapsto \frac {az+b}{cz+d}$, with
$a$, $b$, $c$, $d$ normalized so that $ad-bc=1$, we may
associate\cite{footnote2} with it the matrix
$\begin{pmatrix}
a&b\\c&d\end{pmatrix}$ of determinant 1. The composition
of two M\"obius transformations corresponds to multiplication of the
corresponding matrices, and an inverse transformation corresponds to the
inverse matrix. Thus the boost and rotation are associated with matrices
\begin{equation}
B_x(\phi)=\begin{pmatrix}
\cosh \frac \phi 2&-\sinh \frac \phi 2\\
-\sinh \frac \phi 2 & \cosh \frac \phi 2\end{pmatrix},\ \
R(\theta)=\begin{pmatrix}
e^{i\theta/2}&0\\
0&e^{-i\theta/2}
\end{pmatrix},
\end{equation}
which both have the rather special form,
\begin{equation}
\label{eq:mtdisk}
M(\alpha,\beta)=
\begin{pmatrix}
\alpha&\beta\\
\bar \beta&\bar \alpha
\end{pmatrix},
\end{equation}
for complex numbers $\alpha$ and $\beta$ with $\alpha
\bar \alpha- \beta \bar \beta =1$. Furthermore, for any such matrix
$M(\alpha,\beta)$ with $\theta_1=\arg(\alpha)+\arg(\beta)+\pi$,
$\theta_2=\arg(\alpha)-\arg(\beta)-\pi$, and $\phi$ such that $\cosh \frac
\phi 2=|\alpha|$ and $\sinh \frac \phi 2 =|\beta|$, we have
\begin{equation}
M(\alpha,\beta)=R(\theta_1)B_x(\phi)R(\theta_2).
\end{equation}
Thus
the matrices arising from boosts and rotations generate all matrices of
the form in Eq.~(\ref{eq:mtdisk}).

In fact, the M\"obius transformations associated with matrices of the
form in Eq.~(\ref{eq:mtdisk}) are known to be all the
(orientation-preserving) conformal maps of the Poincar\'e disk to
itself.\cite{footnote4} Because every Lorentz transformation (on
(2+1)-dimensional space) must give rise to a conformal map of the disk,
and every such conformal map arises from a product of two rotations and
a boost $B_x$, then not only do all the conformal maps arise from
Lorentz transformations, but also every Lorentz transformation is a
product of at most two rotations and a boost in the $x$-direction.

\subsection{The Upper-Half Plane Model}\label{S8b}

The transformation associated with $\frac 1 {\sqrt{2}}
\begin{pmatrix}1&i\\i&1\end{pmatrix}$ maps the disk conformally onto the
set of points $z=x+iy$ with $y>0$, and results in the upper half-plane
model. The conformal transformations of this model are the M\"obius
transformations corresponding to $2\times2$ real matrices of determinant
1, that is, to the group $SL_2(\mathbb R)$, since
\begin{equation}
\begin{pmatrix}1&i\\i&1
\end{pmatrix}
\begin{pmatrix}\alpha&\beta\\\bar
\beta & \bar \alpha \end{pmatrix}
\begin{pmatrix}1&i\\i&1\end{pmatrix}^{-1}
\end{equation}
ranges through $SL_2(\mathbb
R)$ as $\alpha,\beta$ range through all complex numbers with $\alpha\bar
\alpha-\beta\bar\beta=1$.

Thus Lorentz transformations (on (2+1)-dimensional space) correspond to
elements of $SL_2(\mathbb R)$, and the action of a
Lorentz transformation on rapidity space is simply the action of the
corresponding M\"obius transformation on the upper half-plane model.

\subsection{Extension to Three Spatial Dimensions and the Spinor
Map}\label{S8c}

Although our discussion has been limited to (2+1)-dimensions for ease of
exposition, all the work carries over in a fairly straightforward way to
(3+1)-dimensions (or more). A higher dimensional paraboloid within the
light cone leads to a conformal model of rapidity space, which is now the
interior of a unit ball. The metric
is given by\cite{Landau2}
\begin{equation}
ds^2=\frac 4{(1-x^2-y^2-z^2)^2}(dx^2+dy^2+dz^2),
\end{equation}
and the geodesics are arcs of circles that intersect the bounding sphere
orthogonally. Within the ball, the surfaces formed by pieces of spheres
centered outside the unit ball which intersect the unit sphere
orthogonally should be though of as ``planar,'' because geodesics remain
inside them. Any of these surfaces can be mapped (by a conformal
transformation of the ball to itself) to a disk bounded by the
equator of the ball. The geometry of such a disk arising from its embedding in the ball is the
same as the geometry developed here for the Poincar\'e disk.

Finally, in addition to the ball model, there is an upper half-space
model composed of points in $\mathbb R^3$ where the third coordinate is
positive. Although points in it cannot be naturally identified by
complex numbers --- it is after all three-dimensional --- they can be
identified with certain quaternions $x+iy+jz$, where $z>0$. The
(orientation-preserving) conformal transformations of this space are
identified with matrices in
$SL_2(\mathbb C)$, where the matrix
$\begin{pmatrix}
a&b\\c&d\end{pmatrix}$ acts
by\cite{footnote1}
\begin{equation}
x+iy+jz\mapsto
(a(x+iy+jz)+b)(c(x+iy+jz)+d)^{-1}.
\end{equation}
The correspondence of Lorentz transformations, which give rise to
conformal transformations of the model, to elements of $SL_2(\mathbb
C)$ is usually called the spinor map.\cite{Naber}

\subsection{More Useful Forms of the Thomas-Wigner Rotation}\label{S8d}

Because we have identified in the matrix $B_x(\phi)$ with a boost of
rapidity $\phi$ in the $x$-direction and the matrix
$R(\theta)$ with a spatial rotation through an angle $\theta$ (see 
Sec.~\ref{S8a}), we can derive a relatively simple equation for the
Thomas-Wigner rotation produced by two successive, non-colinear, boosts.

As is easily proved, a pure boost with rapidity $\phi$ in the direction
of $\theta$ can be obtained by first rotating through $-\theta$, then
applying an $x$-boost of $\phi$, and then rotating back by $\theta$. 
Expressing these three operations with matrices, we have
\begin{equation}
R(\theta)B_x(\phi)R(-\theta)=\begin{pmatrix}
\cosh \frac \phi 2 &
-\sinh \frac \phi 2 e^{i\theta}\\ -\sinh \frac \phi 2 e^{-i\theta}&
\cosh \frac \phi 2 \end{pmatrix}.
\end{equation}
Therefore, as shown in Figs., a boost with a
rapidity of $\phi_1$ in the
$x$-direction, followed by a boost of rapidity $\phi_2$ in the
$\theta=\pi-\alpha_1$ direction, corresponds to
\begin{equation}
R(\theta)B_x(\phi_2)R(-\theta)B_x(\phi_1) =\begin{pmatrix}
\cosh \frac {\phi_2} 2 & -\sinh \frac {\phi_2} 2 e^{i\theta}\\ -\sinh
\frac {\phi_2} 2 e^{-i\theta} & \cosh \frac {\phi_2} 2 \end{pmatrix}
\begin{pmatrix}
\cosh \frac {\phi_1} 2 & -\sinh \frac {\phi_1} 2\\ -\sinh
\frac {\phi_1} 2& \cosh \frac {\phi_1} 2 \end{pmatrix}.
\label{eq:tpprod}
\end{equation}

On the other hand, any Lorentz transformation in the direction of
$\omega_1$ can be expressed as the product of a boost
$R(\omega_1)B_x(\phi_3)R(-\omega_1)$ in the $\omega_1$ direction
followed by a rotation through an angle $\omega_2$. Expressing these operations with matrices, we have
\begin{equation}
R(\omega_2)
\big(R(\omega_1)B_x(\phi_3)R(-\omega_1)\big)\label{eq:tpprod7}.
\end{equation}
In the specific case shown in Figs
and , $\omega_2$ is the Thomas-Wigner rotation angle and
$\omega_1$ is the angle $\alpha_3$. Thus, the product in
Eq.~(\ref{eq:tpprod}) must equal the product in Eq.~(\ref{eq:tpprod7})
which, when expressed in matrix form, is
\begin{equation}
\begin{pmatrix}
e^{i\omega_2/2}&0\\ 0&e^{-i\omega_2/2} \end{pmatrix}
\begin{pmatrix}
\cosh \frac{\phi_3} 2 & -\sinh \frac {\phi_3} 2
e^{i\omega_1}\\ -\sinh \frac{\phi_3} 2 e^{-i\omega_1}& \cosh \frac
{\phi_3} 2
\end{pmatrix}.
\label{eq:tpprod8}
\end{equation}
If we solve for
$\omega_2$ by equating the upper left entries of Eqs.~(\ref{eq:tpprod})
and (\ref{eq:tpprod8}), we find
\begin{subequations}
\begin{align}
\omega_2&=2 \arg
\big(\cosh \frac {\phi_1}2\cosh \frac {\phi_2}2 +\sinh
\frac{\phi_1}2\sinh \frac {\phi_2}2 e^{i\theta} \big )\\ 
\omega_2&=2 \arg\big (1 +\tanh \frac {\phi_1}2\tanh \frac
{\phi_2}2e^{i\theta} \big ). \label{eq:tpprod1}
\end{align}
\end{subequations}

Equation~(\ref{eq:tpprod1}) is an algebraic formula for the
Thomas-Wigner rotation $\omega_2$ resulting from a boost with rapidity
$\phi_1$ in the
$x$-direction followed by a boost with rapidity $\phi_2$ in the
$\theta=\pi-\alpha_1$ direction (as shown in Figs. and
\ref{F8}). Note that Eq.~(\ref{eq:tpprod1}) readily produces the
qualitative results we derived in Sec.~\ref{S7a}. For example, it
shows that the Thomas-Wigner rotation will take on values between $-\pi$
and $\pi$, and will approach its largest value when both velocities are
near $c$ and $\theta$ is near $\pi$. Equation~(\ref{eq:tpprod1}) also
shows that the magnitude of the Thomas-Wigner rotation is the same
regardless of the order in which the boosts $\phi_1$ and $\phi_2$ are
applied.

We end this section by noting that the method used to derive
Eq.~(\ref{eq:tpprod1}) also can be used to find the rapidity $\phi_3$
and the angle $\alpha_3$, and to derive the equations given in
Ref.~\onlinecite{Aravind} for $\tan{(\omega_2/2)}$,
$\cosh{\phi_3}$, and $\tan{\alpha_3}$. For example, if we equate the
real and imaginary parts of the upper left entries of
Eqs.~(\ref{eq:tpprod}) and (\ref{eq:tpprod8}), we find
\begin{equation}
\cos{\dfrac{\omega_2}{2}}\cosh{\dfrac{\phi_2}{2}}=
\cosh{\dfrac{\phi_1}{2}}\cosh{\dfrac{\phi_2}{2}}
+\sinh{\dfrac{\phi_1}{2}}\sinh{\dfrac{\phi_2}{2}}\cos\theta,
\label{former}
\end{equation}
and
\begin{equation}
\sin{\dfrac{\omega_2}{2}}\cosh{\dfrac{\phi_2}{2}}=
\sinh{\dfrac{\phi_1}{2}}\sinh{\dfrac{\phi_2}{2}}\sin{\theta}.
\label{latter}
\end{equation}
If we divide Eq.~(\ref{latter}) by
Eq.~(\ref{former}), we obtain 
\begin{equation}
\tan{\dfrac{\omega_2}{2}}=\
\dfrac{\sinh{(\phi_1/2)}\sinh{(\phi_2/2)}\sin\theta}{\cosh{(\phi_1/2)}\cosh{(\phi_2/2)}+\sinh{(\phi_1/2)}\sinh{(\phi_2/2)}\cos\theta}
\label{A3},
\end{equation}
which is Eq.~(2) in Ref.~\onlinecite{Aravind}. If we divide the
numerator and denominator of Eq.~({\ref{A3}}) by
$\sinh{(\phi_1/2)}\sinh{(\phi_2/2)}$, we obtain the simpler
expression\cite{Ref1}
\begin{equation}
\tan{\dfrac{\omega_2}{2}}=\dfrac{\sin\theta}{\cos\theta + D}\label{A4}.
\end{equation}
The coefficient $D$ can be written as
\begin{align}
D&=\left(\dfrac{\cosh\phi_1/2}{\sinh\phi_1/2}\right)
\left(\dfrac{\cosh\phi_2/2}{\sinh\phi_2/2}\right)\label{A12}\\
&=\left(\dfrac{e^{\phi_1}+1}{e^{\phi_1}-1}\right)
\left(\dfrac{e^{\phi_2}+1}{e^{\phi_2}-1}\right),
\label{A5}
\end{align}
which, from Eq.~(\ref{E10}), is simply a ratio
involving Doppler blueshift factors. Alternatively, if we use
Eqs.~(\ref{A10}), (\ref{A11}), and (\ref{equation2b}) in
Eq.~(\ref{A12}), we see that
\begin{equation}
D=\sqrt{\left(\dfrac{\gamma_1 +1}{\gamma_1
-1}\right)\left(\dfrac{\gamma_2 +1}{\gamma_2 -1}\right)}.\label{A6}
\end{equation}
Equation~(\ref{A4}), together with either Eq.~(\ref{A5}) or
(\ref{A6}), is the simplest expression we have seen for the Thomas-Wigner
rotation angle $\omega_2$.

\section{Summary}
We have presented a self-contained derivation of a
relativistic velocity space called rapidity space. We then
demonstrated how this space can be used to visualize and calculate
various effects resulting from the successive application of non-colinear
Lorentz boosts and the relativistic addition of non-colinear velocities.
In particular, we showed how rapidity space provides a geometric approach
to the Thomas-Wigner rotation and the Thomas precession, and how it
offers both qualitative and quantitative insight into these (and other)
effects.

\section*{Acknowledgements}

We would like to thank Professors Elizabeth Allman and Matthew C\^ot\'e
for their generous and expert help with the figures. We also would like
to thank the referees for bringing several interesting articles to our
attention, and for pointing out that Eq.~(\ref{eq:tpprod1}) can be
rewritten in the simpler form of Eq.~(\ref{A4}).


\begin{thebibliography}{99}
\bibitem{Rhodes} In Sec.~\ref{S8c} we discuss the more general case
of the three-dimensional relativistic velocity space obtained from
four-dimensional spacetime.
\bibitem{ftnote3} In this case, by ``conformal,'' we mean that angle
measurements using a natural metric induced from Minkowski space (in a
manner similar to the one developed in this section) coincide with
those using the metric arising from viewing the surface as embedded in
Euclidean 3-space.
\bibitem{ftnote3a} Because our model involves only
two coordinates, $x_1$ and $x_2$, by ``conformal'' we mean that angle
measurements coincide with those using the Euclidean metric in 2-space.
\bibitem{ftnote4} There are other choices of $A$, $B$ and $C$ that will
solve the three equations. First, if
$A=C=0$, then $B=\pm 1$ and $g=\pm r$. These solutions, however, make
the denominator on the right-hand side of Eq.~(\ref{E16}) vanish.
Second, if $B=0$, then $A$ can be arbitrary as long as
$C=1/(4A)$. In this case, however, if we use any value of $A$ other than
$A=1/2$, we end up with the same model of a disk, but one whose radius
is not equal to one. In this case everything is the same, just scaled 
appropriately.
\bibitem{ftnote12} Although the coordinates in a two- or
three-dimensional velocity space are the components of the velocity, the
coordinates in rapidity space are not components of the rapidity;
rather, they are proportional to the coordinates of the velocity, with a
proportionality factor that varies from point to point. The space is
called rapidity space simply to emphasize that the distance from the
origin to any point in it is the rapidity of that point.
\bibitem{ftnote5} This calculation implicitly assumes that the distance
$s$ is to be evaluated along a straight line. By definition, the
distance between two points is the minimum of all the path integrals
connecting them, which means that distances are evaluated along
geodesics (unless otherwise stated).  Thus, we are assuming that any
geodesic that includes the origin is a straight line. We prove this
assumption in Section~\ref{S6}.
\bibitem{Munoz} One exception is given
in Ref.~\onlinecite{Munoz1} in which the result is derived in the lab
frame.
\bibitem{ftnote15} This fact is easily proved by noting
that any four-sided Euclidean figure can be constructed from two
triangles.
\bibitem{footnote2} There are actually two such matrices associated with
any M\"obius transformation, because the matrices $A$ and $-A$
correspond to the same transformation.
\bibitem{Landau2} This is the exercise from Ref.~\onlinecite{Landau1}
discussed in Sec.~I.
\bibitem{footnote1} Note that because of the non-commutativity of
quaternions, the formula for this M\"obius transformation must be
expressed with the ``denominator'' on the right.
\end{thebibliography}
\end{document}